\documentclass[11pt,a4paper]{article}
\usepackage[a4paper,width=170mm,top=25mm,bottom=25mm]{geometry}
\usepackage{amsmath,amssymb}
\usepackage{mathtools}
\usepackage[affil-it]{authblk}

\usepackage[english]{babel}
\usepackage{siunitx}
\usepackage{graphicx}
\usepackage{rotating}
\usepackage{adjustbox}

\usepackage{setspace}
\usepackage{verbatim}
\usepackage{caption}
\usepackage{wrapfig}
\usepackage{color}
\usepackage{subcaption}
\usepackage{booktabs}
\usepackage{bm}
\usepackage[toc,page]{appendix}
\usepackage[breaklinks=true,colorlinks=true,linkcolor=blue,urlcolor=blue,citecolor=blue]{hyperref}
\usepackage[doi=false,isbn=false,url=false,eprint=false,
    backend=biber, 
    natbib=true,
    style=numeric,
    sorting=none,
]{biblatex}
\title{\textbf{\texttt{py5vec}: a modular Python package for the 5-vector method to search for continuous gravitational waves}}

\author[1]{L. D'Onofrio \thanks{Corresponding author: \texttt{luca.donofrio@na.infn.it}}}
\author[ ]{F. Muciaccia}
\author[2]{L. Mirasola}
\author[3,4]{M. Pitkin}
\author[5]{C. Palomba}
\author[5,6]{P. Leaci}
\author[5]{F. Safai Tehrani}
\author[5,6]{F. Amicucci}
\author[6,7]{L. Silvestri}
\author[5]{L. Pierini}

\affil[1]{INFN, Sezione di Napoli, I-80126 Napoli, Italy}
\affil[2]{Universitat de les Illes Balears, IAC3–IEEC, 07122 Palma, Spain}
\affil[3]{CEDAR Audio Ltd., Cambridge, United Kingdom}
\affil[4]{School of Physics \& Astronomy, University of Glasgow, Glasgow, United Kingdom}
\affil[5]{INFN, Sezione di Roma, I-00185 Roma, Italy}
\affil[6]{Università di Roma La Sapienza, 00185 Roma, Italy}
\affil[7]{INFN, Sezione CNAF, I-40127 Bologna, Italy}

\date{}  

\begin{document}

\maketitle

\vspace{-1cm}

\begin{abstract}
We present \texttt{py5vec}, a Python package for implementing and extending the 5-vector method, used to search for continuous gravitational wave (CW) signals.
We also provide a comprehensive theoretical review of the 5-vector method and extend the relative likelihood formalism by marginalizing over the noise variance, resulting in a more robust Student’s t-likelihood, and over the initial phase to account for pulsar glitches.
\texttt{py5vec} provides a modular architecture that separates data representation, signal demodulation, and statistical inference into independent abstract stages. It supports multiple input data formats and interoperates with existing Python software, providing a bridge between different approaches. For example, using a \texttt{bilby}-based interface, \texttt{py5vec} implements Bayesian parameter estimation within the 5-vector formalism for the first time.
The modular design also allows for making exact multi-level and direct comparisons between other software, such as \texttt{cwinpy} and \texttt{SNAG} in MATLAB.
In \texttt{py5vec}, we implement a multidetector targeted search for known pulsars, validated using LIGO data from the O4a run and hardware injections, demonstrating consistent reconstruction of signal parameters.
This package therefore provides a flexible platform for current targeted searches and for future extensions to other CW search strategies.
\end{abstract}

\section{Introduction}
Continuous gravitational waves (CWs) from rapidly rotating neutron stars represent one of the most challenging and promising targets for ground-based gravitational wave detectors. Unlike transient signals, CWs are weak, long-lived, and nearly monochromatic, requiring highly specialized data analysis techniques to extract them from detector noise \cite{2023LRR....26....3R}. Over the years, a variety of search strategies have been developed, depending on the knowledge of the source parameters (sky position, rotation frequency and orbital parameters); targeted and narrowband searches for known pulsars observed by radio/X-ray/Gamma-ray telescopes, directed searches for sources with only the sky position measured (for instance, neutron stars in supernova remnants), and all-sky surveys for completely unknown objects.

Despite their differences, most CW searches share a common physical structure: a deterministic signal model with slowly varying phase evolution, amplitude modulation induced by the detector response, and a small set of unknown parameters that must be inferred or marginalized over. However, practical implementation of these searches and knowledge of source parameters often leads to pipelines that are optimized for a specific analysis strategy, making it difficult to compare methods or reuse components across different search paradigms (see \cite{WETTE2023102880} for a complete review of algorithms and methods).

Within the class of targeted searches, the 5-vector method provides a compact and efficient frequency-domain formulation of the matched filter, based on the residual amplitude modulation of the signal due to the Earth’s sidereal rotation \cite{2010,2014}. This approach has been extensively used in past analyses (e.g. latest results in \cite{O4a}) and has recently been connected to likelihood-based and Bayesian formulations of CW inference \cite{DOnofrio_2025}. The existing implementation of the 5-vector method is embedded in a software environment, \texttt{SNAG} \cite{Snag}, developed in MATLAB and closely tied to specific data formats, limiting its flexibility and extensibility.

In this work, we present \texttt{py5vec}, a Python package that implements the 5-vector formalism within a modular and framework-independent architecture. First, we provide a comprehensive theoretical review of the state-of-the-art of the 5-vector method and the recent likelihood formulation. In addition, we generalize the likelihood formulation by relaxing the assumption of perfectly known Gaussian noise. By treating the noise variance as a nuisance parameter and marginalizing over it, we derive a Student’s t-likelihood formulation improving robustness against noise misestimation. We also extend the likelihood formalism by marginalizing over the initial phase to account for pulsar glitches.

While the current implementation of \texttt{py5vec} focuses on targeted searches, the design is intentionally agnostic to the specific CW search class. The package is built around a minimal set of abstract data objects and transformations that encode the physical core of CW analyses, rather than a particular end-to-end pipeline.

The guiding principles of \texttt{py5vec} are threefold: (i) separation of data representation, signal demodulation, and statistical inference; (ii) interoperability with existing, well-tested software packages commonly used in the CW community; and (iii) modularity that allows analysis components to be reused, replaced, or extended without modifying the core. Within this framework, the 5-vector method emerges as a specific realization of a more general approach, rather than as a closed and self-contained pipeline. By construction, this architecture naturally lends itself to future extensions. Alternative demodulation strategies, different detection statistics, or extension to different searches can be incorporated while preserving the same underlying data model and interfaces. As such, \texttt{py5vec} is not only a tool for current targeted searches,  but also a flexible platform for the development, validation, and comparison of CW analysis methods across different search regimes.

The current implementation of \texttt{py5vec} enables a complete multidetector targeted search for known pulsar allowing for the first time a Bayesian parameter estimation within the 5-vector formalism. Further developments are planned, including an extended comparison with established pipelines and application to real O4~\cite{LIGOScientific:2025snk} dataset from the LIGO-Virgo-KAGRA detectors~\cite{ligo,virgo, KAGRA:2020tym}. 

The paper is organized as follows. In Section \ref{sec:5vector_overview}, we provide a review of the state-of-art of the 5-vector method, while in Section \ref{sec:extension}, we describe the new extension of the 5-vector likelihood formalism. In Section \ref{sec:architecture}, we describe the guiding principles and the general architecture of \texttt{py5vec}. The validation of the \texttt{py5vec} targeted search considering real data and hardware injections is described in Section \ref{sec:validation}. Conclusion and future prospects are finally presented in Section \ref{sec:end}.

\section{Review of the 5-vector method}\label{sec:overview}\label{sec:5vector_overview}

In this section, we provide a review of the 5-vector method, originally proposed in \cite{2010}. The method has been widely used in the CW community and has been extended to narrowband 
\cite{2014}, semicoherent \cite{semicoh}, directed \cite{Amicucci_2025}, and ensemble \cite{mio} searches. The purpose of this section is to provide a complete guide and a future reference for the 5-vector theory development, also in light of the recent work in \cite{DOnofrio_2025}, which introduces a likelihood for the first time within the 5-vector formalism.

\subsection{General principle}
The 5-vector method is generally defined as a matched filter in the frequency domain. After correcting the detector data for Doppler modulation due to the Earth's motion and for the intrinsic spin-down of the source, the signal becomes nearly monochromatic. The time dependence of the detector response, induced by the Earth's sidereal rotation, modulates the signal amplitude and redistributes the signal power into five discrete frequency components at $\omega_0 + k \Omega_\oplus$ \,, where   $\omega_0$ is the CW angular frequency, $\Omega_\oplus$ is the Earth's sidereal angular frequency and $k = 0,\pm1,\pm2$. According to the considered emission model, $\omega_0$ is proportional to the source rotation angular frequency. In the single-harmonic emission model \cite{cwemission}, the CW angular frequency is exactly twice the source rotation angular frequency.

In the complex formalism, firstly described in \cite{2010}, the expected signal is written as:
\begin{equation}
h(t) = H_0 \left( H_+ \textbf{A}^+ + H_\times \textbf{A}^\times \right)
\cdot \textbf{W}\, e^{j(\omega_0 t + \phi_0)} \,,
\end{equation}
where $H_0$ is the overall signal amplitude, $\phi_0$ is the initial phase, $H_{+/\times}$ are the complex polarization amplitudes depending on the polarization angle $\psi$ and the inclination angle $\iota$ (see details in \ref{app:5vec_formalism}). The amplitude $H_0$ is linked to the standard definition of the amplitude $h_0$ (as defined in \cite{JKS}) by:
\begin{equation}\label{corramp}
 H_0=h_0 \sqrt{\frac{1+6\cos^2\iota + \cos^4 \iota}{4}}\,.
\end{equation}
The 5-vector templates $\textbf{A}^{+/\times}$ encode the detector response to the plus and cross polarizations, and $\textbf{W}=e^{j k \Omega_\oplus t}$ is the sidereal modulation vector. The scalar product between two 5-vectors $\textbf{B}$ and $\textbf{C}$ is defined as $\textbf{B}\cdot\textbf{C}=\sum_{k=-2}^{2} B_k C_k^*$.

The data 5-vector components are proportional to the Fourier transforms of the detector data $x(t)$ at the five signal frequencies:
\begin{equation}\label{Xdata_general}
\textbf{X} = \frac{1}{T_{\rm obs}} \int_0^{T_{\rm obs}} x(t)\,\textbf{W}^* e^{-j\omega_0 t}\,dt \,.
\end{equation}Note the normalization factor $1/T_{\rm obs}$ introduced in \cite{DOnofrio_2025}, which is irrelevant in the single-detector case but becomes important when a network of detectors is considered.

In the absence of a signal and assuming stationary Gaussian noise, the components of $\textbf{X}$ are complex Gaussian random variables with zero mean value and variance proportional to the detector's power spectral density and observation time. In contrast, in the presence of a signal, their mean is shifted by a deterministic contribution proportional to the amplitude $H_0$ \cite{DOnofrio_2025}.

The signal model depends linearly on the two complex amplitudes $H_0 H_{+/\times} e^{j\phi}$, which can therefore be estimated by matched filtering in the 5-vector space through
\begin{equation}\label{mf}
\hat{H}_{+/\times} =
\frac{\textbf{X}\cdot\textbf{A}^{+/\times}}{|\textbf{A}^{+/\times}|^2} \,,
\end{equation}
where $|\textbf{A}^{+/\times}|^2=\textbf{A}^{+/\times}\cdot\textbf{A}^{+/\times}$.

A coherent detection statistic is constructed from the squared modulus of the estimators. Originally in \cite{2010}, the statistic is defined as:
\begin{equation}\label{classicS}
\mathcal{S} = |\textbf{A}^+|^4 |\hat{H}_+|^2
+ |\textbf{A}^\times|^4 |\hat{H}_\times|^2 \,.
\end{equation}
The weights $|\textbf{A}^{+/\times}|^4$ account for the different detector responses to the two polarizations and slightly enhance the detection efficiency when the sensitivity to the plus and cross modes is unbalanced \cite{2010}. Recently, it has been shown that the definition in Eq.\eqref{classicS} corresponds to a weak-signal approximation of the $\mathcal{F}$-statistic \cite{Prix_2025}. 

The standard multidetector extension through the definition of the 5n-vector is described in \cite{2014}. In brief, the 5-vectors from each of the $n$ considered detectors are combined together in an array of $5n$ components and matched with the corresponding template 5n-vectors constructed in the same way. 

\subsection{Maximum likelihood formulation}\label{subsec:5vector_likelihood}

In this section, we briefly summarize the maximum likelihood formulation in the 5-vector formalism for the multidetector case, firstly described in \cite{DOnofrio_2025}.
\\Let us consider a network of $n$ detectors with independent, stationary Gaussian noise, different noise power spectral densities $S_i$, and observation times $T_i$ for the $i$-th detector. 
\\The likelihood ratio for the presence of a signal $\bm h$ in the data $\bm{x}$ can be written as
\begin{equation}\label{eq:likratio}
\ln \Lambda =  (\bm{x}|\bm h) - \frac{1}{2}(\bm h|\bm h)  \,,
\end{equation}
where the multidetector scalar product is defined as
\begin{equation}
    (\bm{a}|\bm{b}) = 2 \sum_{i=1}^n \int_0^{T_i} \frac{a_i(t)b_i^*(t)}{S_i}\,dt \,.
\end{equation}

Using the 5-vector decomposition of the signal, the likelihood depends on the complex polarization amplitudes through
\begin{equation}\label{likelihood_MD}
    \ln \Lambda =
    \sum_{p=+,\times}
    \left[ \Re \left\{
    \lambda_p^*
    \sum_{i=1}^n \gamma_i (\textbf{X}_i \cdot \textbf{A}_i^p) \right\}
    -
    \frac{1}{2} |\lambda_p|^2
    \sum_{i=1}^n \gamma_i |\textbf{A}_i^p|^2
    \right]\,,
\end{equation}
where 
\begin{equation}
\lambda_p = H_0 H_p e^{j\phi_0}\,, \qquad \text{and} \qquad \gamma_i = \frac{T_i}{S_i} \,.
\end{equation}
Maximization with respect to the polarization amplitudes yields the maximum-likelihood estimators
\begin{equation}
\hat{H}_p =
\left(
\sum_{i=1}^n \gamma_i \,\textbf{X}_i \cdot \textbf{A}_i^p
\right)
\left(
\sum_{i=1}^n \gamma_i\,|\textbf{A}_i^p|^2
\right)^{-1},
\end{equation}
leading to a coherent detection statistic formally equivalent to the $\mathcal{F}$-statistic. Indeed, as shown in \cite{DOnofrio_2025}, ``weighting'' the data 5n-vectors as
\begin{equation}\label{5nvec_ML}
\Tilde{\textbf{X}}=\left[\sqrt{\gamma_1}\, \textbf{X}_1,...,\sqrt{\gamma_n}\, \textbf{X}_n\right] 
\end{equation}
and the template 5n-vectors as
\begin{equation}\label{A5nvec_ML}
\Tilde{\textbf{A}}^{+/\times}=\left[\sqrt{\gamma_1}\, \textbf{A}^{+/\times}_1,...,\sqrt{\gamma_n}\, \textbf{A}^{+/\times}_n\right]\,,
\end{equation}
 the maximum likelihood statistics is:
\begin{equation}\label{lambdamax}
    \mathcal{S}_\mathcal{F}=|\Tilde{\textbf{A}}^+|^2 |\hat{H}_+|^2 + |\Tilde{\textbf{A}}^\times|^2 |\hat{H}_\times|^2\,.
\end{equation}

The maximization of the likelihood yields coefficients of the detection statistic proportional to $|\textbf{A}^{+/\times}|^2$. This choice is statistically equivalent to the $\mathcal{F}$-statistic, which has been originally derived from a time-domain matched filter \cite{JKS}. The $\mathcal{S}_\mathcal{F}$ definition entails a noise distribution that is independent of the considered pulsar, as well as an analytical signal distribution under the assumption of Gaussian noise.

\subsection{Bayesian inference}
The likelihood formulation above provides a natural starting point for Bayesian inference. Given the set of signal parameters
\begin{equation}
    \bm{\theta} = \{H_0, \phi_0, \psi, \iota\} \,,
\end{equation}
the posterior probability density is
\begin{equation}
    p(\bm{\theta}|\textbf{X}) \propto
    \mathcal{L}(\textbf{X}|\bm{\theta}) \,\pi(\bm{\theta}) \,,
\end{equation}
where $\mathcal{L}(\textbf{X}|\bm{\theta})$ is the full likelihood and $\pi(\bm{\theta})$ denotes the prior distribution.

The likelihood depends quadratically on the complex polarization amplitudes, a direct consequence of assuming Gaussian noise in the data. The inference is performed by directly sampling from the posterior probability distribution of the physical signal parameters $\{H_0, \phi, \psi, \iota\}$, similar to what has been recently proposed in \cite{Ashok}.

In the multidetector case, the Bayesian framework automatically incorporates different noise levels and observation times through the weights $T_i/S_i$, ensuring optimal combination of different datasets. This makes the 5-vector likelihood formalism particularly well suited for joint fast inference across detector networks, Bayesian inference, and hierarchical analyses (as in \cite{ensbayes}).

\section{Extension of the 5-vector likelihood formalism}\label{sec:extension}
The likelihood formulation described in Section~\ref{subsec:5vector_likelihood} relies on two key assumptions: stationary Gaussian noise with known variance, and a phase model that accurately describes the signal over the entire observation time.

In this section, we extend the 5-vector likelihood formalism in two directions. First, by marginalizing over the noise variance, we derive a Student’s t-likelihood that relaxes the assumption of perfectly known Gaussian noise. Second, we generalize the signal model to incorporate phase discontinuities induced by pulsar glitches, which are sudden increases in the rotation frequency observed in the timing of many radio pulsars \cite{PhysRevD.96.063004}.

\subsection{5-vector Student's t-likelihood}\label{subsec:studentT}

The likelihood ratio introduced in Eq.~\eqref{eq:lnL_unmarg} is derived under the assumption of stationary Gaussian noise with known power spectral density. In the multidetector formulation, this information is incorporated through the weights $\gamma_i$ which define the weighted 5$n$-vectors in Eqs.~\eqref{5nvec_ML}–\eqref{A5nvec_ML}. This weighting is equivalent to whitening the data with respect to the noise model, so that the Gaussian likelihood ratio is written in a space where the noise variance is effectively unitary.

In practice, however, the detector noise levels are only estimated and may deviate from the ideal whitening assumption due to spectral estimation errors, residual non-stationarity, or calibration uncertainties. To account for this effect, the Gaussian likelihood can be generalized by introducing an unknown global noise variance $S$ in the weighted space.

Let us consider the residual weighted 5$n$-vector
\begin{equation}
    \Tilde{\textbf{R}} =
    \Tilde{\textbf{X}} - \Tilde{\bm h}(\bm{\theta}),
\end{equation}
where $\Tilde{\textbf{X}}=\{\Tilde{\textbf{X}}_1,\dots,\Tilde{\textbf{X}}_n \}$ and $\Tilde{\bm h}$ denote the signal model
\begin{equation}
    \Tilde{\bm h} = \{ \Tilde{\bm h}_1,\dots  \Tilde{\bm h}_n\} \qquad \text{with} \qquad \Tilde{\bm h}_i = \sum_{p=+,\times} \lambda_p \, \Tilde{\textbf{A}}^p_i \,.
\end{equation} 
Under the Gaussian noise assumption, the full likelihood can be written as
\begin{equation}
    \mathcal{L}(\Tilde{\textbf{X}}|\bm{\theta},S)
    \propto
    S^{-D}
    \exp\left[
    -\frac{1}{S}
    (\Tilde{\textbf{R}}|\Tilde{\textbf{R}})
    \right],
\end{equation}
where $D=5n$ is the number of complex degrees of freedom and $(\cdot|\cdot)$ denotes the scalar product in the weighted space.

The likelihood ratio of Eq.~\eqref{eq:lnL_unmarg} is recovered from the expression above by fixing $S=1$ and removing the data-only term $(\Tilde{\textbf{X}}|\Tilde{\textbf{X}})$.
Introducing $S$ therefore amounts to relaxing the assumption of perfect noise whitening.

Treating $S$ as a nuisance parameter and adopting a scale invariant prior $p(S)\propto 1/ S$ as in \cite{bayesian}, the marginal likelihood becomes
\begin{equation}
    \mathcal{L}(\Tilde{\textbf{X}}|\bm{\theta})
    \propto
    \int_0^\infty
    S^{-D-1}
    \exp\left[-\frac{(\Tilde{\textbf{R}}|\Tilde{\textbf{R}})}{S}
    \right] dS \,.
\end{equation}
The integral can be solved analytically, yielding
\begin{equation}
    p(\Tilde{\textbf{X}}|\bm{\theta})
    \propto
    (\Tilde{\textbf{R}}|\Tilde{\textbf{R}})^{-D},
\end{equation}
which corresponds to a multivariate Student's t-likelihood. The associated log-likelihood is therefore
\begin{equation}
    \ln \mathcal{L}(\Tilde{\textbf{X}}|\bm{\theta})
    =
    - D \ln (\Tilde{\textbf{R}}|\Tilde{\textbf{R}}) + \mathrm{const}\,.
\end{equation}
and in the 5-vector formalism,
\begin{equation}(\Tilde{\textbf{R}}|\Tilde{\textbf{R}}) = 
    \sum_{i=1}^n |\Tilde{\textbf{X}}_i|^2 -
    2\Re\left\{\sum_{i=1}^n \sum_{p=+,\times} 
    \lambda_p (\Tilde{\textbf{X}}_i \cdot  
    \Tilde{\textbf{A}}_i^p)  \right\}+  
    \sum_{i=1}^n  
    \sum_{p=+,\times} 
    |\lambda_p|^2
    |\Tilde{\textbf{A}}_i^p|^2 \,.
\end{equation}
The log-likelihood ratio is
\begin{equation}\label{tlikelihood}
    \ln \Lambda_t
    \equiv
    \ln \frac{(\Tilde{\textbf{R}}|\Tilde{\textbf{R}})^{-D}}{(\Tilde{\textbf{X}} |  \Tilde{\textbf{X}})^{-D}} = - D \left( \ln ( \Tilde{\textbf{R}}|\Tilde{\textbf{R}}) - \ln (\Tilde{\textbf{X}} |  \Tilde{\textbf{X}}) \right) \,.
\end{equation}
Marginalizing over $S$ effectively replaces the exponential dependence on the residual power with a power-law dependence, producing heavier tails.
As a consequence, the Student's t-likelihood is more robust to noise misestimation and outliers, and yields broader and more conservative posteriors.

\subsection{5-vector $\phi_0$-marginalized likelihood}
\label{subsec:glitching_pulsars}

For glitching pulsars, the phase evolution of the CW signal is disrupted at each glitch epoch. As a consequence, phase coherence between different inter-glitch intervals cannot be maintained in general, and the initial phase should be treated as an independent nuisance parameter for each segment. So far, the 5-vector method analyzed each inter-glitch period independently and then summed the corresponding detection statistic (e.g. as in \cite{O4a}). In \texttt{py5vec}, we adopt an incoherent approach, analyzing each inter-glitch interval separately and combining the information at the likelihood level. 

For a given inter-glitch interval (assuming a single detector analysis for simplicity), the log-likelihood ratio is written as
\begin{equation}
\ln \Lambda
=
\gamma
\sum_{p=+,\times}
\left[
\lambda_p^* \, (\textbf{X}\cdot\textbf{A}^p)
+
\lambda_p \, (\textbf{X}\cdot\textbf{A}^p)^*
-
|\lambda_p|^2 \, | \textbf{A}^p |^2
\right],
\label{eq:lnL_unmarg}
\end{equation}
that is Eq.~\eqref{likelihood_MD} for $n=1$. Defining
\begin{equation}
\mathcal{Z} \equiv \sum_{p=+,\times} H_p^* (\textbf{X}\cdot\textbf{A}^p),
\end{equation}
Eq.~(\ref{eq:lnL_unmarg}) can be rewritten as
\begin{equation}
\ln \Lambda
=
\gamma
\left[
2 H_0 \Re\left( e^{-j\phi_0} \mathcal{Z} \right)
-
H_0^2
\sum_{p=+,\times} |H_p|^2 | \textbf{A}^p |^2
\right].
\end{equation}
As in the case of the $\mathcal{F}$-statistic likelihood \cite{marg_phi0,Fbayesnew}, asssuming a uniform prior,
\begin{equation}
p(\phi_0) = \frac{1}{2\pi}, \qquad \phi_0 \in [0,2\pi),
\end{equation}
the likelihood can be marginalised analytically over $\phi_0$:
\begin{equation}
\Lambda_{\phi_0}
=
\int_0^{2\pi} \frac{d\phi_0}{2\pi} \,
\exp(\ln \Lambda).
\end{equation}
This yields
\begin{equation}
\Lambda_{\phi_0}
=
\exp\left(
-\gamma H_0^2
\sum_{p=+,\times} |H_p|^2 | \textbf{A}^p |^2
\right)
I_0\left(
2\gamma H_0 |\mathcal{Z}|
\right),
\end{equation}
where $I_0$ is the modified Bessel function of the first kind and order zero. The corresponding log-likelihood is therefore
\begin{equation}
\ln \Lambda_{\phi_0}
=
\ln I_0\left(2\gamma H_0 |\mathcal{Z}|\right)
-
\gamma H_0^2
\sum_{p=+,\times} |H_p|^2 | \textbf{A}^p |^2.
\label{eq:lnL_marg}
\end{equation}

For a glitching pulsar with multiple inter-glitch intervals, and under the assumption that the initial phases are independent between intervals, the total likelihood is obtained as the product of the marginalised likelihoods for each segment, or equivalently, as the sum of the corresponding log-likelihoods. 

\section{\texttt{py5vec} architecture}\label{sec:architecture}

The architecture of \texttt{py5vec} has been explicitly designed to be independent of the input data framework and to separate data manipulation from statistical inference into independent  composable layers. This represents a conceptual departure from the existing implementation of the 5-vector pipeline within the MATLAB-based \texttt{SNAG} software.

At the lowest level, \texttt{py5vec} supports multiple heterogeneous input formats, including Band Sampled Data (BSD) \cite{ornella} files, long Fourier-transformed data (\textit{Muciaccia et al.} in preparation), and already heterodyned time series \cite{bayesian} produced by \texttt{cwinpy} \cite{Pitkin2022}. Each format is handled by a dedicated loader that maps the original data to a minimal abstract object,  a \texttt{BandlimitedComplexDataTimeseries}, with associated metadata. This many-to-one mapping decouples the analysis pipeline from any specific data framework and allows different data products to be directly compared.

A key novelty of the approach lies in the explicit treatment of signal demodulation as an abstract transformation acting on the data object. Corrections for Doppler modulation and intrinsic spin-down, using, for example, heterodyne strategies, are implemented as interchangeable modules with identical interfaces. In practice, this enables different demodulation schemes such as SNAG-style heterodyned data or phase reconstruction based on \texttt{cwinpy} to be \emph{hot-swapped} within the same pipeline. This design choice makes it possible to directly and exactly compare different demodulation procedures, which are often difficult to isolate in traditional implementations.

Once a \texttt{CorrectedBandlimitedComplexDataTimeseries} is initialized, detector response functions are computed through \texttt{LAL.ComputeDetAMResponseSeries} from \texttt{LALSuite}~\cite{lalsuite, swiglal} (see \ref{app:equivalence_5vec_Fstat}, for details on the computation and comparisons between the different formalisms) and using the \texttt{cwinpy.PulsarParameter} class to read the pulsar ephemerides. In \texttt{py5vec}, the data and template 5-vectors are constructed in a unified manner, independent of the upstream data representation or demodulation strategy. The 5-vector formalism thus emerges as a well-defined intermediate layer connecting the physical signal model to the statistical analysis.

At the final stage, \texttt{py5vec} allows different statistical paradigms to operate on the same 5-vector objects. In addition to the traditional frequentist approach \cite{2010}, the package supports Bayesian inference through a direct interface with \texttt{bilby} \cite{bilby_paper} for the first time within the 5-vector formalism. This architecture naturally accommodates the recent developments described in \cite{DOnofrio_2025} and summarized in Section \ref{subsec:5vector_likelihood} that reinterpret the 5-vector statistic in terms of likelihood-based and Bayesian approaches.

\begin{figure}
  \centering
  \includegraphics[
      scale=0.62,
      trim=0cm 1cm 0cm 1cm,
      clip
    ]{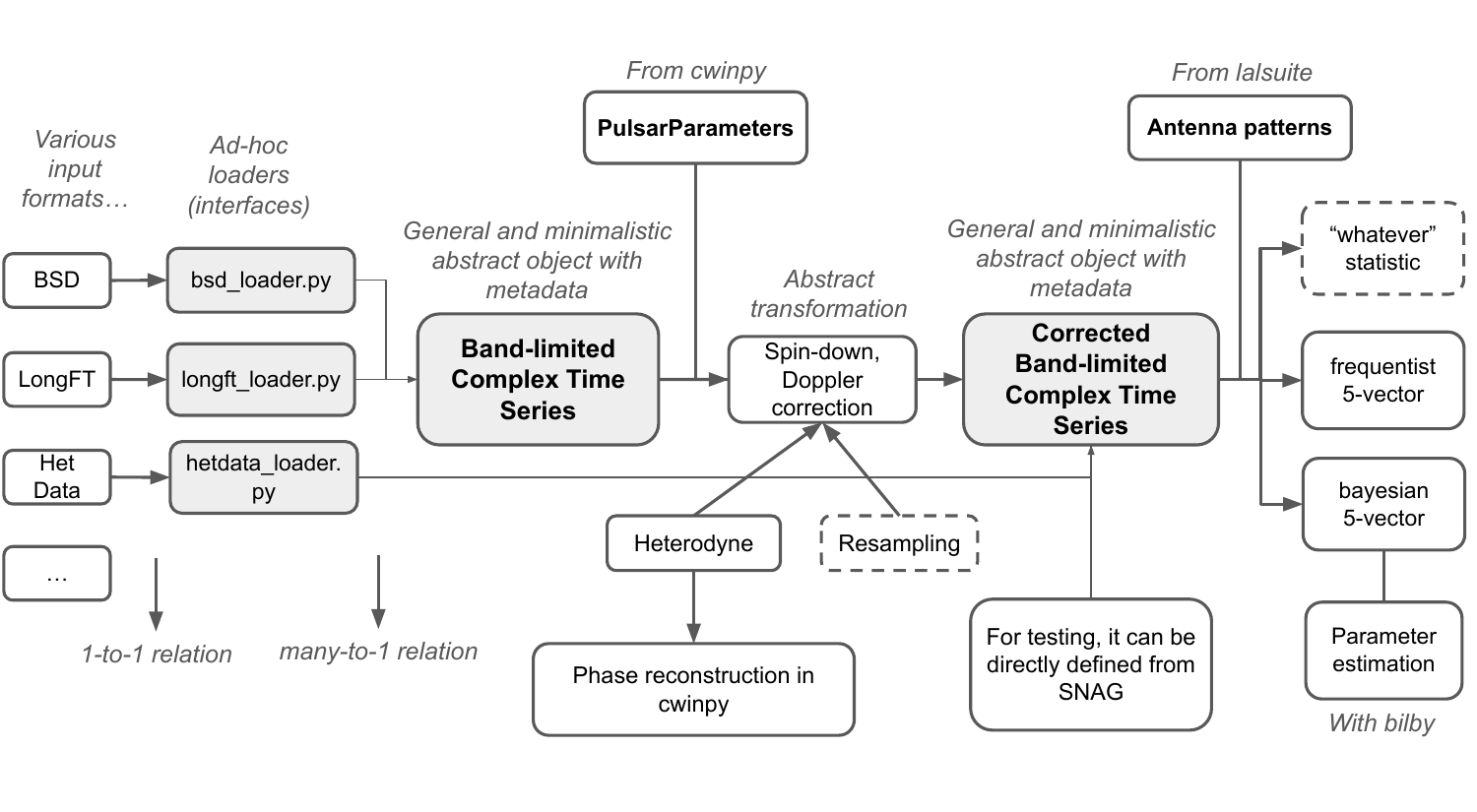}
  \caption{Overall architecture of the \texttt{py5vec} framework. Dashed boxes are still to be implemented but they are shown for how they could be integrated in the future. }
  \label{fig:architecture-overview}
\end{figure}

Overall, the modular and layered design of \texttt{py5vec} is a software engineering choice, but also a methodological one. A clear separation between data handling, signal demodulation, and statistical inference, as shown in Figure \ref{fig:architecture-overview}, facilitates validation studies and the exploration of alternative analysis strategies. This flexibility is essential for future applications to different searches and systematic comparison between different methods.

\subsection{Loading and correcting data}
We developed a Python module, \texttt{bsd\_loader}, to read BSD files \cite{2019} distributed in MATLAB .mat format. The loader extracts the BSD complex time series together with the associated time and frequency metadata using \texttt{scipy} \cite{2020SciPy-NMeth}. It identifies data gaps from both the BSD gap flags metadata and missing or null samples in the time series. From this information, it builds an explicit gap mask and computes the effective observation time used in the subsequent analysis.

The initialized \texttt{BandlimitedComplexDataTimeseries} is then corrected by applying the heterodyne demodulation, namely by multiplying the data by the exponential factor $e^{-j \Phi_{\mathrm{corr}}(t)}$.
This procedure corrects the pulsar signal for phase modulations induced by spin-down, Doppler, and relativistic effects, effectively unwinding the apparent phase evolution of the source. In \texttt{py5vec}, the phase evolution $\Phi_{\mathrm{corr}}(t)$ of the expected signal is reconstructed using the \texttt{cwinpy} class \texttt{HeterodynedCWSimulator}.

We also implemented a dedicated \texttt{hetdata\_loader} module to read \texttt{cwinpy} \texttt{HeterodynedData} products stored in HDF5 format. This data structure is also used as input to the time-domain matched filter ($\mathcal{F}$-statistic) pipeline \cite{O4a,JKS}. The loader reads the heterodyned complex time series and applies preprocessing steps within \texttt{cwinpy}, including outlier removal and zero padding onto a uniform sampling grid. The original and padded sampling frequencies are estimated from the timestamp differences, and the zero-padded series is rescaled by the corresponding normalization factor to preserve spectral consistency with the \texttt{cwinpy} power-spectrum convention. The module then builds a \texttt{CorrectedBandlimitedComplexDataTimeseries} object with explicit time and frequency intervals. A boolean gap mask is also generated directly from the padded data values, providing an explicit representation of missing samples for downstream analyses.

\subsection{5-vector computation}
The data 5-vector is constructed by projecting the corrected time series onto the five sidereal frequency components using an explicit discrete Fourier transform, similarly to the \texttt{SNAG} implementation. Given a complex time series $x(t_k)$ sampled with step $\Delta t$ over an observation time $T_{\rm obs}=N \Delta t$, the $k$-th component of the 5-vector is computed as
\begin{equation}
A_k = \frac{\Delta t}{T_{\rm obs}}
\sum_{n=0}^{N-1} x(t_n)
e^{- j 2\pi f_k t_n},
\quad
f_k = \frac{\omega_0 + k \Omega_\oplus}{2\pi} \,.
\end{equation}
This corresponds to a discrete-time approximation of the Fourier integral, with explicit normalization by $T_{\rm obs}$, while $t_0$ is the actual start time of the considered data segment. Using absolute timestamps in the phase factor preserves the correct phase alignment across data sets with different start times and naturally supports the multidetector extension without introducing an additional reference-time shift parameter. The noise properties of the data 5-vector are estimated directly from the data by constructing ensembles of 5-vectors from off-source frequency bins, excluding the sidereal harmonics associated with the signal.

The presence of gaps in the data is handled by distinguishing between the nominal observation time and the effective observation time. The effective observation time $T_{\rm eff}$ is computed from the number of valid (non-zero) samples in the corrected time series. This quantity enters both the noise variance estimate and the normalization of the detection statistic.
In practice, the normalization factor in Eq.\eqref{eq:lnL_unmarg} (and in the same way, for the weights in Eq.\eqref{5nvec_ML} and Eq.\eqref{A5nvec_ML}) is defined as
\begin{equation}
\gamma = \frac{T_{\rm obs}^2}{S_{\Delta f} T_{\rm eff}}\,.
\end{equation}
The sensitivity $S_{\Delta f}$ is computed from the variance of the band-limited timeseries, or from the variance of the data 5-vector distribution evaluated on off-source frequencies (see Figure \ref{fig:X_dist}). Under the assumption of stationary Gaussian noise, the two procedures are statistically equivalent.

Template 5-vectors are computed from antenna pattern functions for the plus and cross polarizations. The antenna responses are generated using the detector model provided by \texttt{LALSuite}~\cite{lalsuite, swiglal} and are sampled at the same times as the data. Template 5-vectors can be obtained by applying the same Fourier-based projection used for the data 5-vector. The gap pattern inferred from the loaded data is consistently applied to the antenna responses, ensuring that data and templates 5-vectors are affected in an identical way. An analytic implementation of the 5-vector templates (following \cite{2010}) is also provided and used as an internal cross-check. 

The noise distribution of the data 5-vector is used to reconstruct the noise distribution of the polarization amplitude estimators and of the associated detection statistics. Based on these ingredients, frequentist estimators for the signal parameters are implemented as described in \cite{2010}. To assess the detection significance, the $p$-value is computed either directly from the empirical noise distribution and, in case of stationary Gaussian noise, compared with the analytic computation.

\subsection{Bayesian parameter estimation}
\label{subsec:Bayesian_pe}

The Bayesian parameter estimation targets the four signal parameters
$\{ H_0, \phi_0, \psi, \iota \}$. The likelihood ratio is implemented by defining a dedicated \texttt{bilby.Likelihood} class, which constructs either the Gaussian likelihood in Eq.\eqref{likelihood_MD}, the Student's t-likelihood in Eq.\eqref{tlikelihood} and the $\phi_0$-marginalized likelihood in Eq.\eqref{eq:lnL_marg}. For data segments separated by glitches, the total likelihood is constructed as an incoherent sum of independent phase-marginalized likelihoods, each corresponding to a contiguous inter-glitch interval.

The priors on the signal parameters are assumed to be uniform within physically motivated bounds (following the convention in \texttt{cwinpy} \cite{Jones2015ParameterCA,O4a}): the overall amplitude is taken as $H_0 \in [0, 10^{-21}]$, the initial phase as $\phi_0 \in [0, 2\pi)$, the polarization angle as $\psi \in [0, \pi/2)$, and the inclination parameter as $\cos\iota \in [-1, 1]$. For sources where independent measurements of the orientation or polarization angles are available, informative priors are adopted instead; in these cases, each prior is modeled as a Gaussian distribution centered on the measured value with the corresponding standard deviation.

Posterior sampling is performed using the \texttt{dynesty}~\cite{Speagle:2019ivv} nested sampler as implemented in \texttt{bilby}. From the resulting posterior distributions, point estimates, credible intervals, and Bayesian upper limits on the signal amplitude $H_0$ are derived. Posterior projections and correlations among parameters are visualized using corner plots. The upper limit on $H_0$ is converted to an upper limit on $h_0$ averaging Eq.\eqref{corramp} over the unknown parameter $\cos \iota$ and obtaining a conversion factor of $\approx 1.32$.

\begin{figure}[t!]
    \centering
    \begin{subfigure}{\textwidth}
        \centering
        \includegraphics[scale=0.32, trim=2cm 0cm 0cm 0cm,
      clip]{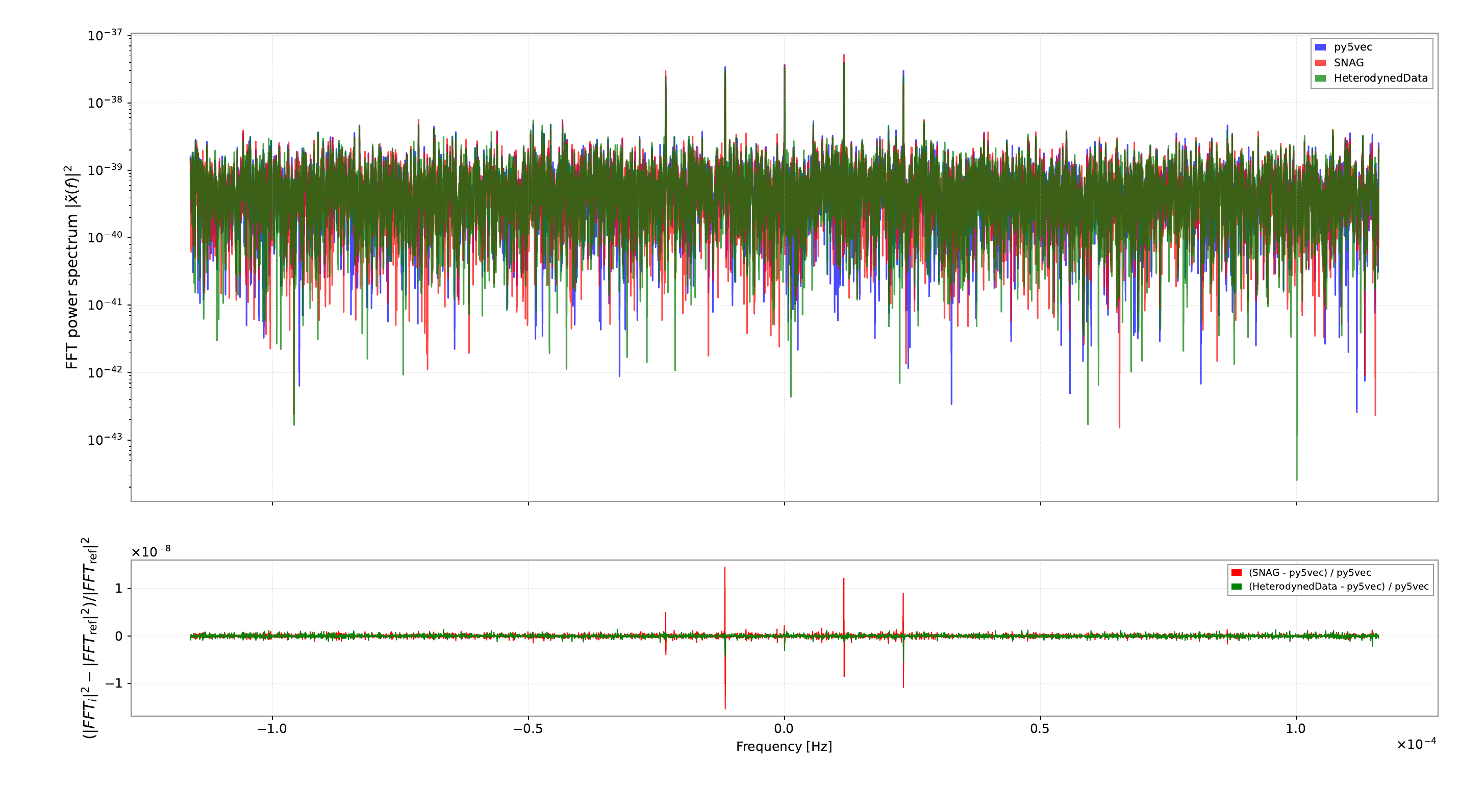}
        \caption{}
    \end{subfigure}

    \vspace{-35pt}

    \begin{subfigure}{\textwidth}
        \centering
        \includegraphics[scale=0.32, trim=2cm 0cm 0cm 0cm,
      clip]{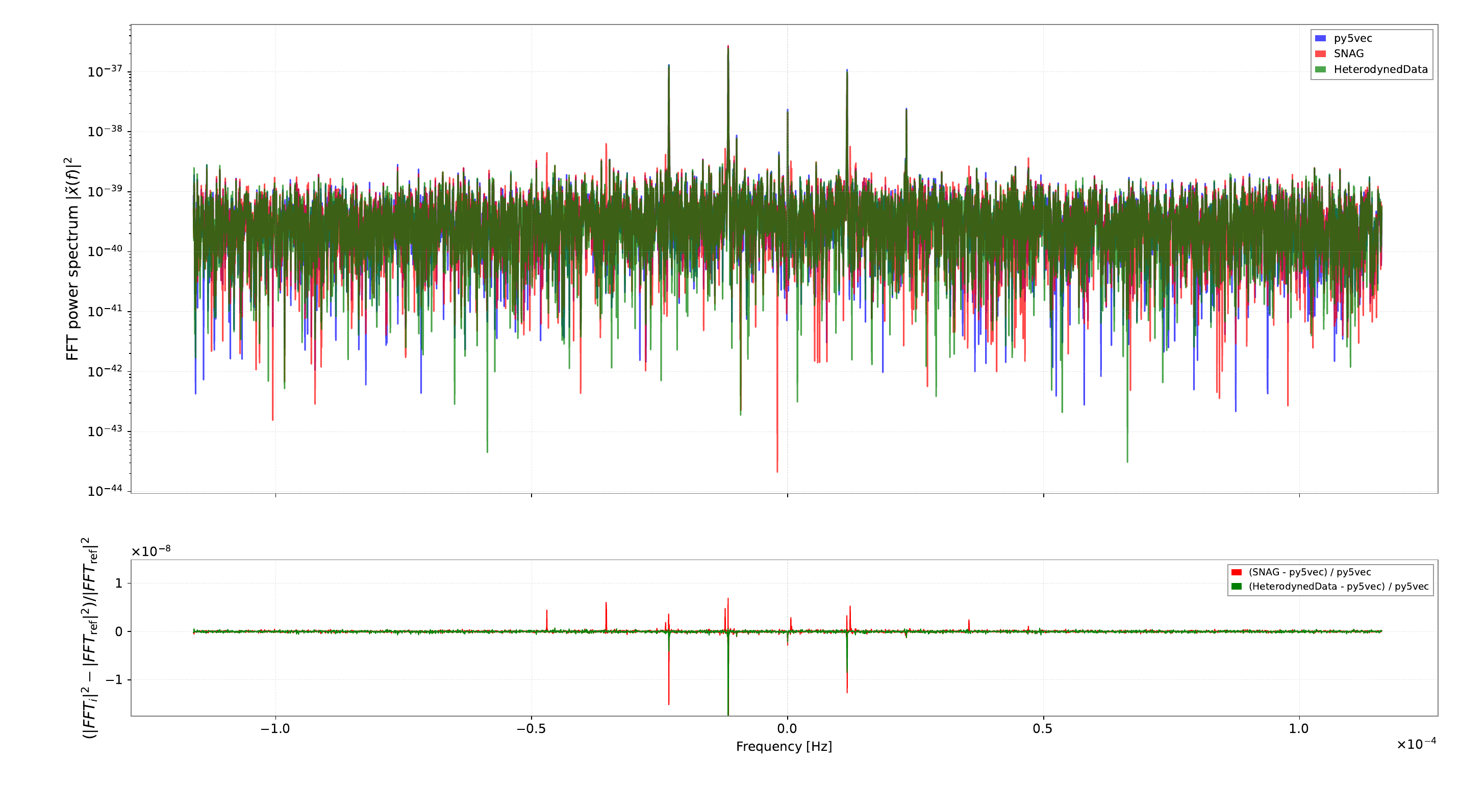}
        \caption{}
    \end{subfigure}
    \vspace{-35pt}
    \caption{
    Comparison between the squared amplitude of the fast Fourier transform of the heterodyned data from \texttt{py5vec} (input BSD data, phase reconstructed using \texttt{cwinpy}), \texttt{SNAG} (input BSD data, phase reconstructed using independent MATLAB routine in \texttt{SNAG}), \texttt{HeterodynedData} (input format used by \texttt{cwinpy}).
    The top plot shows a small frequency band of the band-limited data for the hardware injection HI3, considering the LIGO Livingston O4a dataset, while the bottom plot displays the same comparison for HI16 and the LIGO Hanford O4a detaset. For each plot, a small panel shows the fractional residuals normalized to \texttt{py5vec}.     }
    \label{fig:comparison}
\end{figure}

\section{Validation with hardware injections}\label{sec:validation}
In this section, we describe a set of tests of the implementation of the targeted search in \texttt{py5vec} and comparisons with well-established pipelines (such as \texttt{SNAG} and \texttt{cwinpy}), using real data from the O4a run \cite{LIGOScientific:2025snk, O4} for the two LIGO detectors, as well as hardware injections. Hardware injections \cite{HI, Baxi:2026wbr} are simulated signals physically added to the detectors’ data by displacing the detectors’ test masses, mimicking the detectors’ response to a particular CW signal. The monitoring of the hardware injections during the O4 observing run is described in \cite{O4HI}.

\subsection{Heterodyned data comparison}

The modular architecture of \texttt{py5vec} enables a direct comparison of intermediate analysis products generated by independent pipelines. In particular, heterodyned data can be compared even if they originate from different data frameworks, preprocessing strategies, and phase evolution reconstruction procedures.

Figure~\ref{fig:comparison} shows the squared amplitude of the fast Fourier transform, computed without windowing and using a coherence time equal to the full O4a dataset, for the hardware injections HI3 (top-plot) and HI16 (bottom-plot), which mimics a pulsar in a binary system. Three different heterodyned data object are compared coming from independent pipelines: the \texttt{py5vec} heterodyned data, obtained loading BSD data and reconstructing the phase evolution via \texttt{cwinpy} (blue line), the \texttt{SNAG} heterodyned data, obtained loading BSD data with an independent phase reconstruction in MATLAB (red line), and the fully independent \texttt{HeterodynedData} product from \texttt{cwinpy} (green line). Despite sharing the same BSD input format, the \texttt{SNAG} and \texttt{py5vec} pipelines implement completely independent phase reconstruction procedures, while the \texttt{HeterodynedData} differs in both data handling and phase generation. 

The spectra in Figure~\ref{fig:comparison} show an excellent overlap. The fractional residuals normalized to \texttt{py5vec} are shown in the bottom panels of each plot. Away from the five signal peaks, the residuals fluctuate around zero. In the bins corresponding to the signal peaks, the residuals exhibit a systematic offset, indicating that \texttt{py5vec} reconstructs slightly more signal power than \texttt{cwinpy} (green line). For the \texttt{SNAG} heterodyned data (red line), some effects are clearly convolved with the signal peaks, producing additional small oscillations that appear as almost nine peaks in the residual plot for HI16. The origin of these very small deviations is not yet fully understood.
Overall, the figure demonstrates that all three heterodyned data are consistent, with only very minor deviations around the signal peaks that require a more systematic analysis. The comparison of the data/template 5-vectors computed from these pipelines is described in Table \ref{tab:comp_5vec}.

So far, pipeline comparisons have been performed only on analysis products, such as upper limits or parameter estimation for hardware injections. The heterodyned data comparison therefore provides a stringent validation of consistency at the time series level rather than at the level of final inference products. This is the first time such a comparison has been presented and it demonstrates a very good agreement among the different inputs and phase reconstructions, as expected. 

\subsection{Noise distributions}
\begin{figure}[t]
\centering
        \includegraphics[scale=0.4]{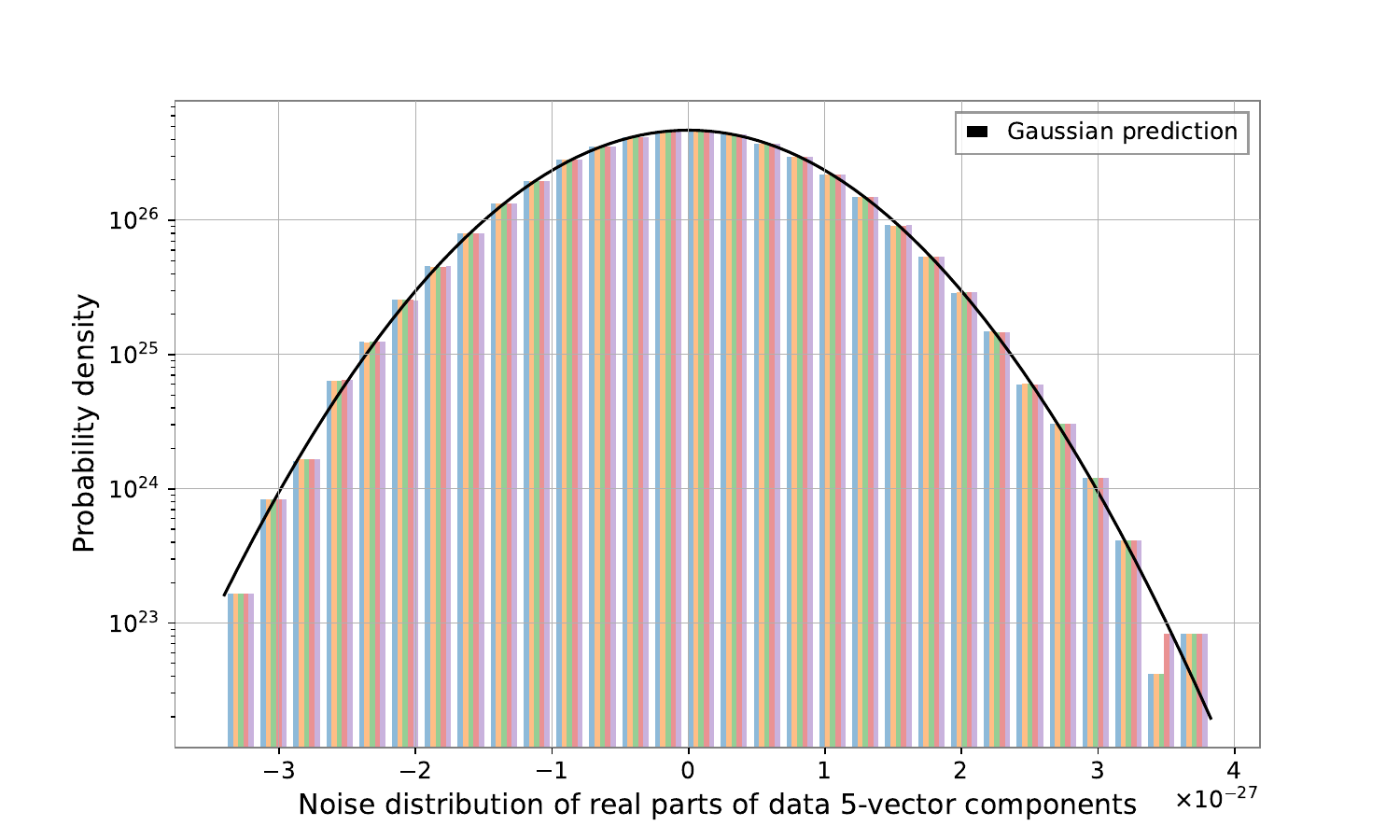}
    \caption{Distribution of each component of the data 5-vectors computed at off-source frequencies to reconstruct from the data the noise distribution of the random variables defined. The black line is the theoretical estimation assuming Gaussian noise in the frequency band considered ($[108,109]$ Hz). For reference, the distribution is inferred from the analysis of the hardware injection HI3 in O4a data from the LIGO Livingston detector.}
    \vspace{-10pt}
    \label{fig:X_dist}
\end{figure}

The noise properties of the data 5-vector are estimated directly from the data by constructing ensembles of 5-vectors from off-source frequency bins, excluding the sidereal harmonics associated with the signal. These empirical noise distributions are compared with the analytic distribution inferred in the assumption of stationary Gaussian noise. Figure \ref{fig:X_dist} shows the distribution of the off-source 5-vectors for the real part of each component, inferred from the analysis of HI3 in the LIGO Livingston O4a dataset. 

\begin{figure}[t]
\centering
\begin{subfigure}{0.49\textwidth}
    \centering
    \includegraphics[width=\linewidth]{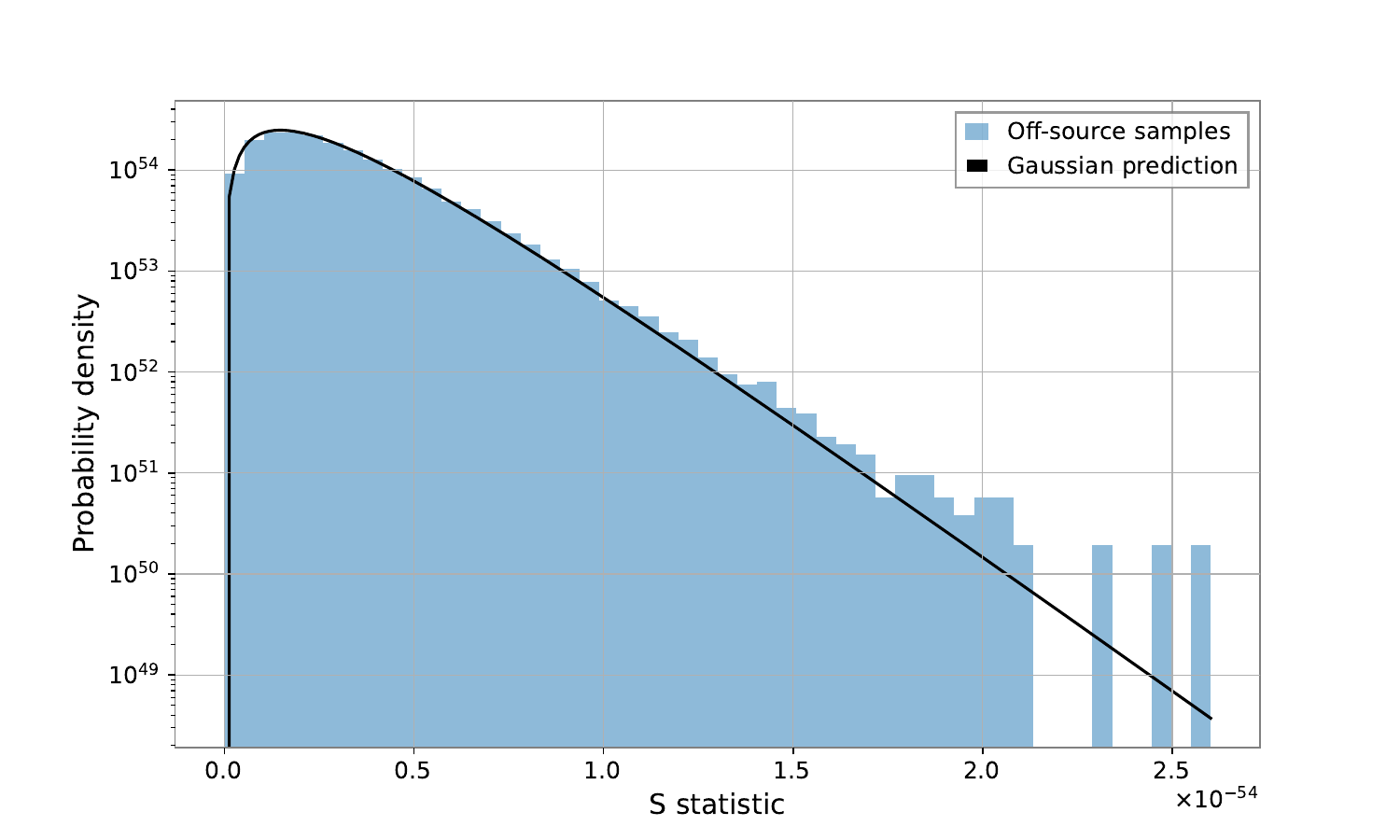}
\end{subfigure}
\hfill
\begin{subfigure}{0.49\textwidth}
    \centering
    \includegraphics[width=\linewidth]{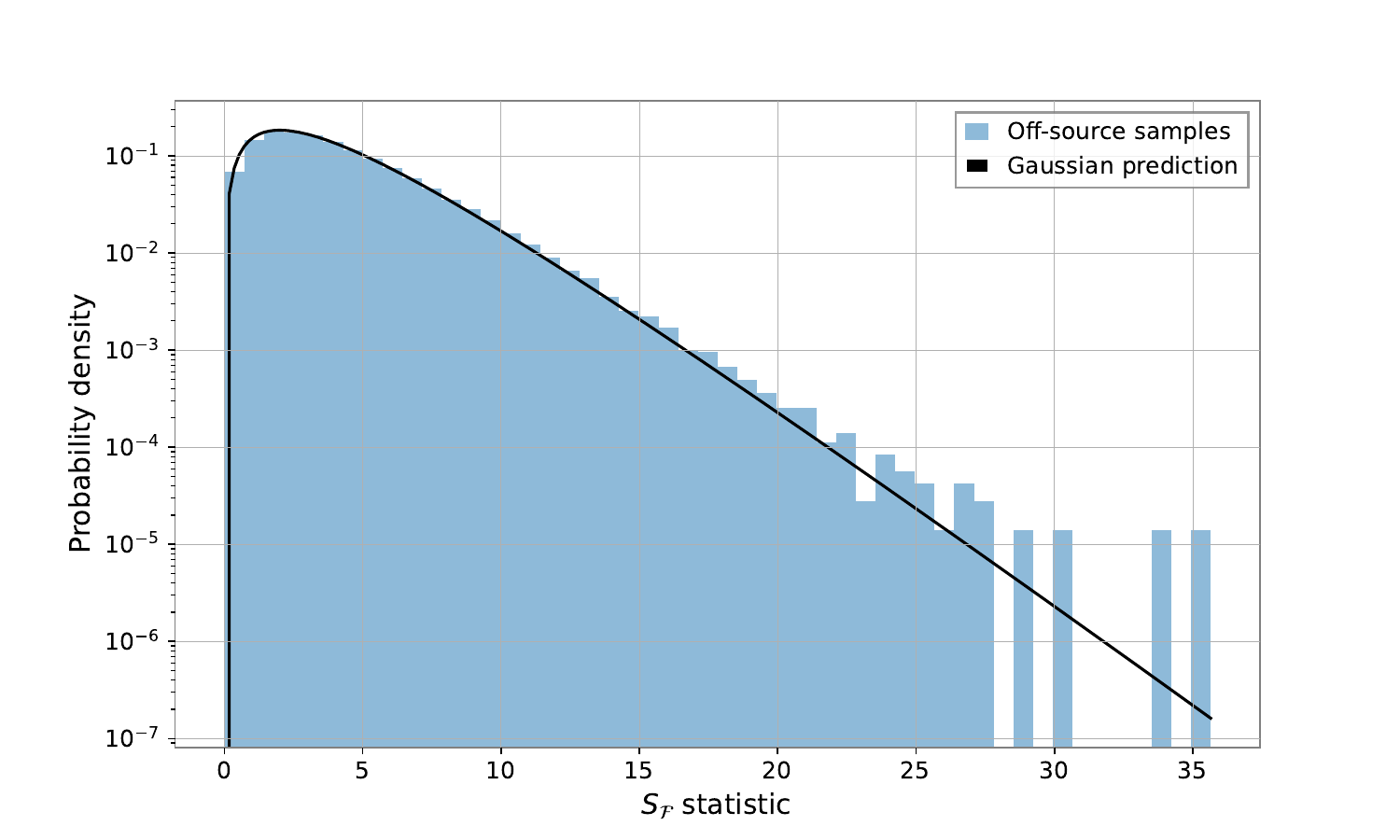}
\end{subfigure}
\caption{Noise distribution of the detection statistics defined in Eq.~\eqref{classicS} (left) and Eq.~\eqref{lambdamax} (right). The black lines correspond to the theoretical noise distributions assuming Gaussian noise in the considered frequency band ($[108,109]$ Hz). For reference, the distributions are inferred from the analysis of the hardware injection HI3 in O4a data from the LIGO Livingston detector.}
\vspace{-10pt}
\label{fig:Sstat_Fstat_dist}
\end{figure}

The off-source 5-vectors are used to derive the expected distributions of the matched filter estimators and of the detection statistic (Figure \ref{fig:Sstat_Fstat_dist} for $\mathcal{S}_\mathcal{F}$ in Eq.\eqref{lambdamax}, and for $\mathcal{S}$ in Eq.\eqref{classicS}). The experimental distributions for HI3 (selected BSD frequency band $[108, 109]$ Hz) agree well with the expected analytic distributions, showing that in this case the O4a noise is stationary and Gaussian for the selected frequency band.

The multidetector analysis is implemented by extending the 5-vector formalism to a coherent combination of multiple interferometers.
For a given set of detectors, the data and template 5-vectors are combined through detector-wise weighted
scalar products following Eq.\eqref{5nvec_ML}-\eqref{A5nvec_ML}. The off-source frequencies strategy is used to construct the expected noise distributions of the data 5-vector for each detector and randomly combined to construct the 5n-vectors and the corresponding distribution.

\subsection{Hardware injections analysis}
In this section, we show the Bayesian  results of the \texttt{py5vec} targeted search from the multidetector analysis of HI3 and HI16 in O4a data. 
\begin{figure}[t]
\centering
        \includegraphics[scale=0.43,
      trim=1cm 0cm 0cm 0cm,
      clip]{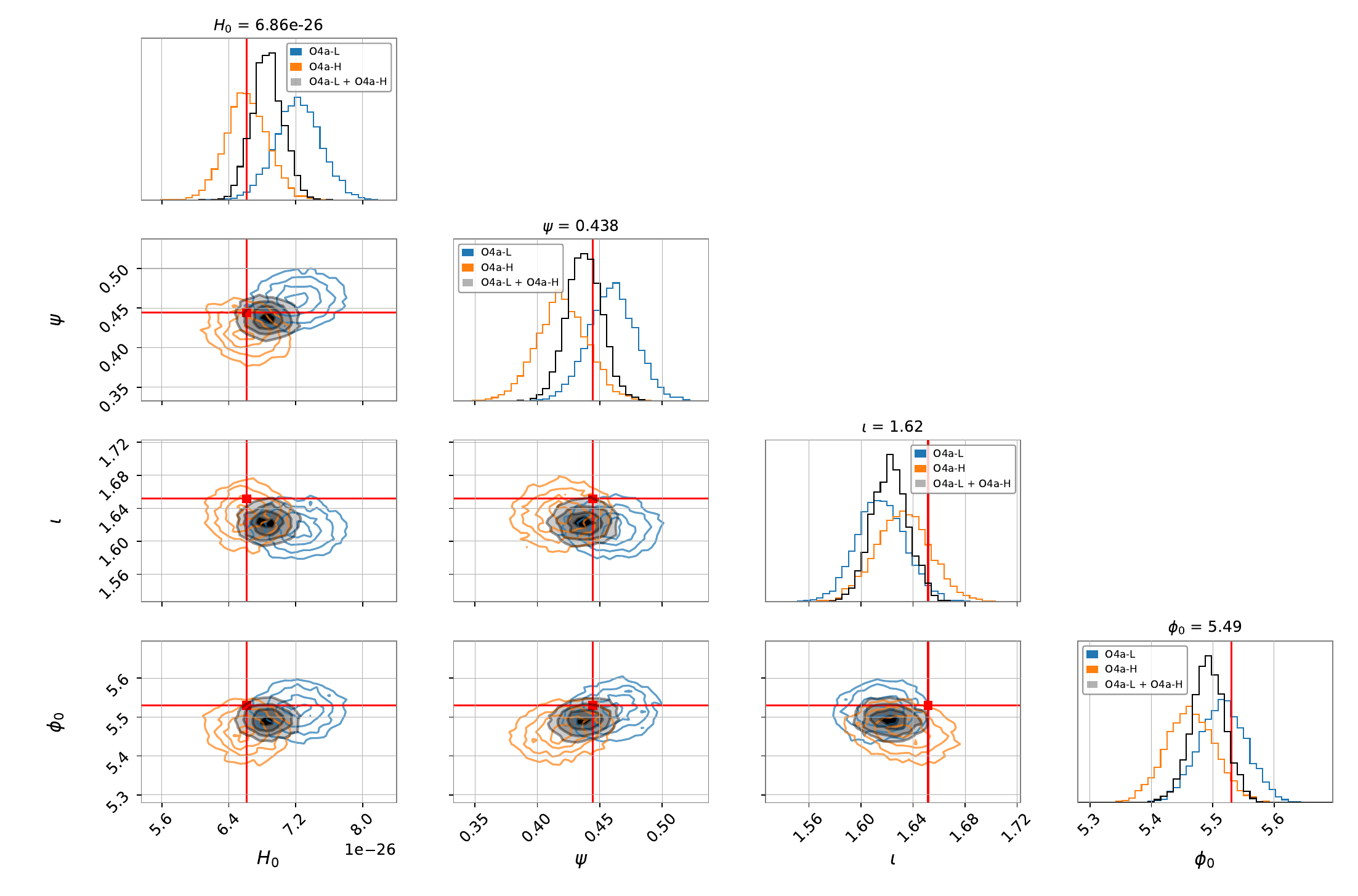}
    \caption{Posterior distribution functions for the signal parameters from the analysis of the hardware injection HI3 considering the O4a dataset of the two LIGO detectors (L1 is LIGO Livingston, H1 is LIGO Hanford) and the analysis of single and multidetector. The red dots/lines show the injected value.}
    \vspace{-10pt}
    \label{fig:HI3_O4a_PE}
\end{figure}
Figure \ref{fig:HI3_O4a_PE} shows the posterior distributions of the signal parameters obtained for HI3.
The results of the individual LIGO detectors (L1 for LIGO Livingston and H1 for LIGO Hanford) are compared with the coherent multidetector analysis considering the Gaussian likelihood in Eq.\eqref{likelihood_MD}.
For each parameter, the one-dimensional marginalized posteriors are shown on the diagonal, while the off-diagonal panels display the corresponding two-dimensional joint distributions with contours corresponding to the 1-, 2-, and 3-$\sigma$ credible regions.
The multidetector posterior is consistently narrower than the single-detector ones, reflecting the increased information content of the coherent combination. The evidence (logarithm of the Bayes factor) further quantifies this improvement, with values of  $\approx 344$ for LIGO Livingston, $\approx 320$ for LIGO Hanford, and $\approx 676$ for the multidetector analysis, indicating a significantly stronger support for the signal when combining the detectors coherently, as expected.
The injected parameter values are indicated by red vertical lines in the one-dimensional plots and by red markers in the two-dimensional projections, and are found to be well contained within the high-posterior-density regions. 

\begin{figure}[t]
\centering
        \includegraphics[scale=0.43,
      trim=1cm 0cm 0cm 0cm,
      clip]{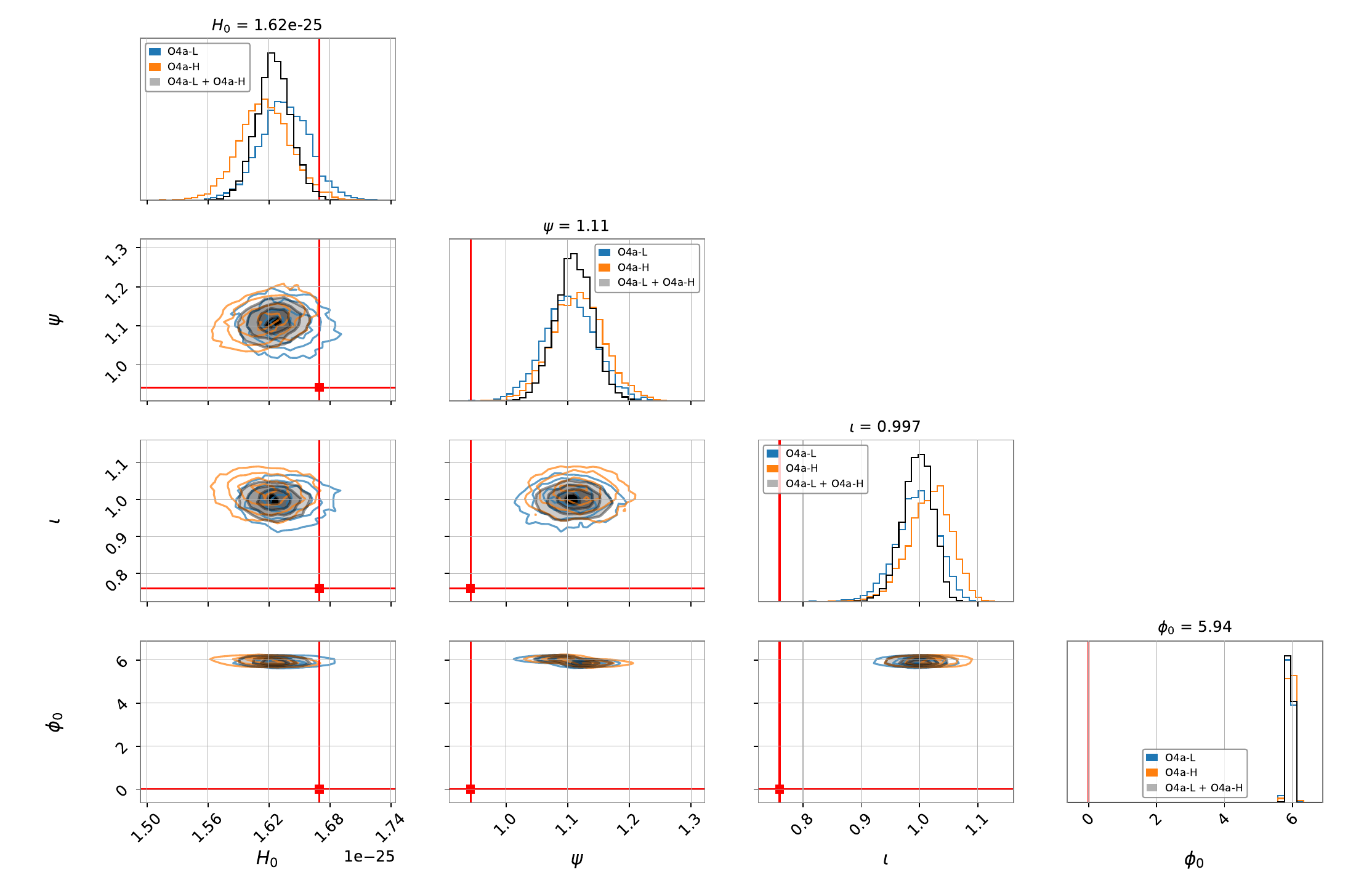}
    \caption{Posterior distribution functions for the signal parameters from the analysis of the hardware injection HI16 (binary system) considering the O4a dataset of the two LIGO detectors (L1 is LIGO Livingston, H1 is LIGO Hanford, MD refers to the multidetector result) and the analysis of single and multidetector. The red dots/lines show the injected value. For this plot, we use the Gaussian  likelihood defined in Eq.\eqref{likelihood_MD}.}
    \vspace{-10pt}
    \label{fig:HI16_O4a_PE}
\end{figure}

Figure \ref{fig:HI16_O4a_PE} shows the results of the Bayesian parameter estimation for HI16 which mimics a CW signal from a binary system, considering the Gaussian likelihood in Eq.\eqref{likelihood_MD}. 
HI16 has a value of $\iota \approx \pi/4$ that entails a value of $\eta \approx 1$ (see Figure \ref{fig:eta}), close to the circular polarization condition. If $\eta \approx 1$, the polarization amplitudes $\lambda_p \propto e^{j(\phi_0 - 2\psi)}$ and there is a strong $\psi - \phi_0$ degeneracy (see Table \ref{tab:deg}). The posterior on $\phi_0$ peaks at almost $2\pi$ while the injected value is $0$. The $\psi - \phi_0$ degeneracy is not visible in Figure \ref{fig:HI16_O4a_PE}, probably due to the high SNR of the injection combined with the Gaussian likelihood. 

\begin{figure}[t]
\centering
        \includegraphics[scale=0.43,
      trim=1cm 0cm 0cm 0cm,
      clip]{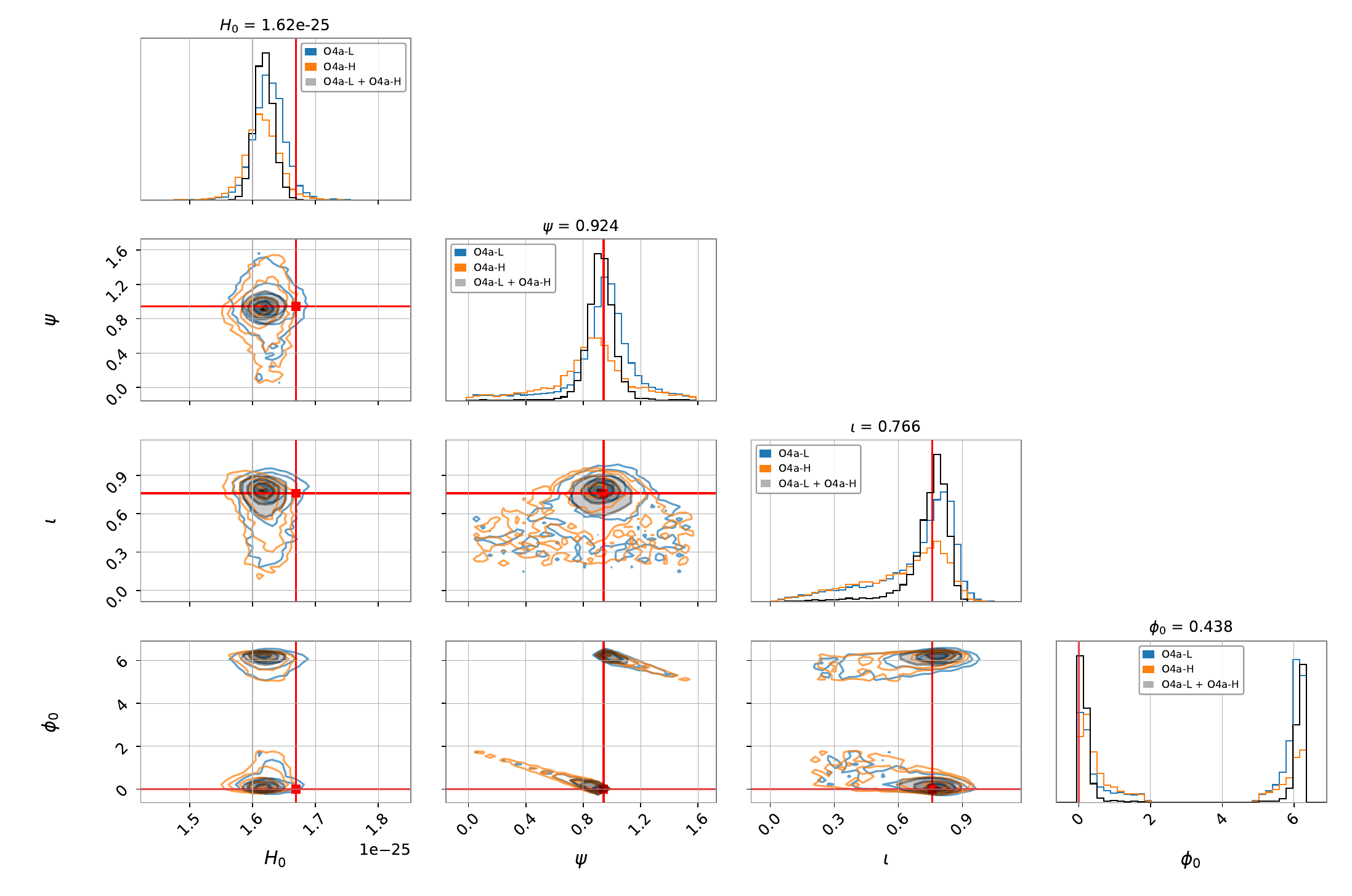}
    \caption{Posterior distribution functions for the signal parameters from the analysis of the hardware injection HI16 (binary system) considering the O4a dataset of the two LIGO detectors (L1 is LIGO Livingston, H1 is LIGO Hanford, MD refers to the multidetector result) and the analysis of single and multidetector. The red dots/lines show the injected value. For this plot, we use the Student's t-likelihood defined in Eq.\eqref{tlikelihood}.}
    \vspace{-10pt}
    \label{fig:HI16_O4a_PE_ts}
\end{figure}

For the same datasets and injection, Figure \ref{fig:HI16_O4a_PE_ts} is inferred instead from the Student's t-likelihood defined in Eq.\eqref{tlikelihood}. The degeneracy is fully observed and the posteriors are in very good agreement with the \texttt{cwinpy} results in Figure \ref{fig:cwinpy_HI16_O4a_PE}, which assumes a similar Student's t-likelihood \cite{bayesian}. The difference in the amplitude posterior is due to the formalism; $H_0$ is proportional to $h_0$ by the factor in $\cos \iota$ (see Eq.\eqref{corramp}) that masks the strong correlation seen in Figure \ref{fig:cwinpy_HI16_O4a_PE}.

Computationally, the analysis is very efficient. A single likelihood evaluation takes \( \sim 8.6 \times 10^{-5} \) s, and the nested sampling run (with $\text{nlive} = 1024$) for a single detector completes in less than two minutes. All runs were performed on a 13th Gen Intel i7-1355U CPU (12 cores/threads), demonstrating that the method is easily feasible on a standard desktop PC.

\section{Conclusion and future prospects}\label{sec:end}

The modular design of \texttt{py5vec} naturally enables a broad range of future developments, both in terms of scientific applications and methodological extensions. While the current implementation focuses on targeted CW searches, the underlying architecture has been conceived to support more general classes of CW analyses without modifying the physical core of the framework.

A primary short-term goal is the application of \texttt{py5vec} to a full targeted search on O4 data, including extensive validation on real interferometric data. Ongoing efforts also include a systematic comparison with established pipelines (as \texttt{cwinpy} and the $\mathcal{F}$-statistic pipeline) to quantify the impact of different data representations, demodulation strategies, and statistical estimators on both detection efficiency and parameter estimation.

Beyond targeted searches, the framework can be extended to more complex signal models. In particular, narrowband, semicoherent, and directed searches can be accommodated by adapting the parameter space exploration while preserving the same abstract data objects and interfaces.

The explicit separation between demodulation and statistical inference also opens the possibility of incorporating alternative analysis strategies, such as resampling-based approaches \cite{Amicucci_2025} or more agnostic searches to account for frequency wandering of the signal \cite{spinwand}.  

Finally, the Python-based nature of the framework facilitates integration with external tools for data pre-processing (such as the folding procedure \cite{folding}), follow-up studies \cite{Mirasola}, population analyses \cite{mio2, mio3}, and parameter estimation using the $\mathcal{F}$-statistic likelihood \cite{Ashok}. Interfaces to existing packages, such as \texttt{pyFstat}~\cite{Keitel:2021xeq, PhysRevD.97.103020} or custom Bayesian samplers, can be developed to provide independent validation and cross-checks. 

In the longer term, \texttt{py5vec} aims to serve not only as an analysis tool but also as a flexible platform for method development, testing, and simulation within the CW community. In particular, the modular representation of data and detector responses makes it a natural platform for signal simulations and methodological studies in the context of future observatories such as the Einstein Telescope.

\section*{Acknowledgments}

We gratefully acknowledge helpful discussions with colleagues in the LIGO–Virgo–KAGRA Continuous Waves Working Group. 

This material is based upon work supported by NSF's LIGO Laboratory, which is a major facility fully funded by the National Science Foundation.

This work was supported by the Universitat de les Illes Balears (UIB)  with funds from the Programa de Foment de la Recerca i la Innovació de la UIB 2024-2026 (supported by the yearly plan of the Tourist Stay Tax ITS2023-086); the Spanish Agencia Estatal de Investigación grants PID2022-138626NB-I00, RED2024-153978-E, RED2024-153735-E, funded by MICIU/AEI/10.13039/501100011033 and the ERDF/EU; and the Comunitat Autònoma de les Illes Balears through the Conselleria d'Educació i Universitats with funds from the ERDF (SINCO2022/18146-Plataforma HiTech-IAC3-BIO) and by COST action SCALES CA24139, supported by COST (European Cooperation in Science and Technology). 

\appendix

\section{Equivalence between the 5-vector and $\mathcal{F}$-statistic implementation}
\label{app:equivalence_5vec_Fstat}

The 5-vector formalism and the standard $\mathcal{F}$-statistic approach describe the same detector response to the incoming CW signal, but differ in the way the time dependence of the detector response is parametrized. In this section, we explicitly show their equivalence by comparing the antenna pattern functions and the associated phase conventions.

\subsection{From the $\mathcal{F}$-statistic formalism to the 5-vector formalism}\label{app:5vec_formalism}

We start from the standard expression of the CW signal in \cite{eterodina},
\begin{equation}
h(t)=\frac{1}{2} F_{+}(t;\psi)\,h_0(1+\cos^2\iota)\cos\Phi(t)
      + F_{\times}(t;\psi)\,h_0\cos\iota\,\sin\Phi(t)\,,
\end{equation}
where $F_{+}$ and $F_{\times}$ are the detector antenna pattern functions \cite{JKS},
$\iota$ is the inclination angle with respect to the line-of-sight, $\psi$ the polarization angle of the signal (determined by the position angle of
the spin axis, projected on the sky),
and $\Phi(t)$ the signal phase that encodes the initial phase $\phi_0$.

Introducing the complex representation of the phase,
\(
\cos\Phi(t)=\Re\{e^{j\Phi(t)}\},\;
\sin\Phi(t)=\Re\{-i\,e^{j\Phi(t)}\},
\)
the strain can be written as
\begin{equation}
h(t)=\Re\left\{
\left[\frac{h_0}{2}(1+\cos^2\iota)F_{+}(t;\psi)
      - j\,h_0\cos\iota\,F_{\times}(t;\psi)
\right] e^{j\Phi(t)}
\right\}.
\end{equation}

The antenna patterns are expanded in terms of the sidereal response
functions $a(t)$ and $b(t)$ \cite{JKS}:
\begin{align}\label{eq:F+Fx}
F_{+}(t;\psi) &= a(t)\cos2\psi + b(t)\sin2\psi, \\
F_{\times}(t;\psi) &= b(t)\cos2\psi - a(t)\sin2\psi.
\end{align}
It is convenient to factor out a common amplitude scale by defining
\begin{equation}
\eta \equiv -\frac{2\cos\iota}{1+\cos^2\iota},
\qquad
H_0 \equiv \frac{h_0}{2}(1+\cos^2\iota)\sqrt{1+\eta^2}.
\end{equation}
Let us define the normalized polarization amplitudes as
\begin{equation}
 H_{+} = \frac{\cos2\psi - j\,\eta\sin2\psi}{\sqrt{1+\eta^2}}, \qquad
 H_{\times} = \frac{\sin2\psi + j\,\eta\cos2\psi}{\sqrt{1+\eta^2}},
\end{equation}
which satisfy the normalization condition
\begin{equation}
| H_{+}|^2 + | H_{\times}|^2 = 1.
\end{equation}
Substituting and grouping the terms proportional to $a(t)$ and $b(t)$ gives
\begin{equation}
h(t)=\Re\left\{
\left[a(t)\,H_{+}(\psi,\iota) + b(t)\,H_{\times}(\psi,\iota)\right]
e^{j\Phi(t)}
\right\}\,.
\end{equation}
Defining the template functions
\begin{equation}
A^+(t) \equiv a(t),
\qquad
A^\times(t) \equiv b(t),
\end{equation}
one finally obtains the form of the 5-vector representation:
\begin{equation}
h(t)=\Re\left\{
H_0\left[A^{+}(t) H_{+}(\psi,\iota)
       + A^{\times}(t) H_{\times}(\psi,\iota)\right]
e^{j\Phi(t)}
\right\}.
\end{equation}
In this expression, the time dependence is entirely contained in the known sidereal functions $a(t)$ and $b(t)$ (hence in $A^{+/\times}(t)$), whose Fourier decomposition produces the five discrete frequency components that define the 5-vector formalism, while the source parameters are encoded in the polarization amplitudes.

\subsection{Antenna patterns}

In the $\mathcal{F}$-statistic framework \cite{JKS}, the detector response to the plus and cross polarizations is described in the functions $F_{+,\times}(t,\psi)$. The functions $a(t)$ and $b(t)$ in Eq.\eqref{eq:F+Fx} are periodic functions on the hour angle \cite{JKS}:
\begin{align}
a(t) &= a_0 + a_{1c}\cos H(t) + a_{1s}\sin H(t)
      + a_{2c}\cos 2H(t) + a_{2s}\sin 2H(t) \,, \\
b(t) &= b_{1c}\cos H(t) + b_{1s}\sin H(t)
      + b_{2c}\cos 2H(t) + b_{2s}\sin 2H(t) \,,
\end{align}where the coefficients $a_{kc}, a_{ks}, b_{kc}, b_{ks}$ depend only on the detector geometry. The hour angle $H(t)$ of the source is defined as
\begin{equation}
H(t) = \mathrm{LST}(t) - \alpha = \mathrm{GST}(t) + \beta - \alpha \,.
\end{equation} where $\alpha$ be the right ascension of the source and $\beta$ the detector longitude, $\mathrm{GST}(t)$ is the Greenwich Mean Sidereal Time and $\mathrm{LST}(t)$ is the Local Sidereal Time. The sidereal rotation of Earth introduces a modulation at the sidereal angular frequency $\Omega_\oplus$, such that
\begin{equation}
\mathrm{GST}(t) = \Omega_\oplus t + \mathrm{GST}(t_0) \,.
\end{equation}
In \cite{2010}, the expressions for the template functions $A^{+/\times}(t)$ in the 5-vector formalism are
\begin{equation}
    \begin{aligned}
        A^+(t) &= a_{0}
+ a_{1c}\cos \Omega_\oplus t
+ a_{1s}\sin \Omega_\oplus t
+ a_{2c}\cos 2\Omega_\oplus t
+ a_{2s}\sin 2\Omega_\oplus t,
\\
A^\times(t) &=
b_{1c}\cos \Omega_\oplus t
+ b_{1s}\sin \Omega_\oplus t
+ b_{2c}\cos 2\Omega_\oplus t
+ b_{2s}\sin 2\Omega_\oplus t,
    \end{aligned}
\end{equation}
For consistency, the time $\Omega_\oplus t$ corresponds to
\begin{equation}
    \Omega_\oplus t \rightarrow
\mathrm{GST}(t_0)
+ \Omega_\oplus dt \cdot (0:N)
- \alpha
+ \beta.
\end{equation}
Indeed, in \cite{2010}, the CW signal is expressed in terms of the template 5-vectors $\textbf{A}^{+/\times}$,
\begin{equation}
    h(t)=H_0\textbf{A}\cdot \textbf{W}e^{j\omega_0 t}\equiv H_0 \left( H_+\textbf{A}^+ + H_\times\textbf{A}^\times\right)\cdot \textbf{W} e^{j\omega_0 t}
\end{equation}
where $\textbf{W}=e^{jk\Omega_\oplus t},\,k=0,\pm1,\pm2$ while the expressions of $\textbf{A}^{+/\times}$ are \cite{2010}
\begin{equation}
    \begin{aligned}
A^{+}_{-2} &= \left(\frac{a_{2c}}{2} + j\,\frac{a_{2s}}{2}\right) e^{j2(\alpha-\beta)}
&
A^{\times}_{-2} &= \left(\frac{b_{2c}}{2} + j\,\frac{b_{2s}}{2}\right) e^{j2(\alpha-\beta)}
\\
A^{+}_{-1} &= \left(\frac{a_{1c}}{2} + j\,\frac{a_{1s}}{2}\right) e^{j(\alpha-\beta)}
&
A^{\times}_{-1} &= \left(\frac{b_{1c}}{2} + j\,\frac{b_{1s}}{2}\right) e^{j(\alpha-\beta)}
\\
A^{+}_{0} &= a_{0}
&
A^{\times}_{0} &= 0
\\
A^{+}_{1} &= \left(\frac{a_{1c}}{2} - j\,\frac{a_{1s}}{2}\right) e^{-j(\alpha-\beta)}
&
A^{\times}_{1} &= \left(\frac{b_{1c}}{2} - j\,\frac{b_{1s}}{2}\right) e^{-j(\alpha-\beta)}
\\
A^{+}_{2} &= \left(\frac{a_{2c}}{2} - j\,\frac{a_{2s}}{2}\right) e^{-j2(\alpha-\beta)}
&
A^{\times}_{2} &= \left(\frac{b_{2c}}{2} - j\,\frac{b_{2s}}{2}\right) e^{-j2(\alpha-\beta)}
\end{aligned}
\end{equation}

Using the relation between hour angle and sidereal time,
\begin{equation}
H(t) = \Omega_\oplus t + \beta - \alpha + \mathrm{GST}(t_0) \,,
\end{equation}
the trigonometric functions appearing in $a(t)$ and $b(t)$ can be rewritten as complex exponentials:
\begin{equation}
\cos(k H(t)) \pm j \sin(k H(t)) = 
e^{\pm j k (\Omega_\oplus t + \beta - \alpha +\mathrm{GST}(t_0))} \,.
\end{equation}
The 5-vector components are obtained as the discrete Fourier coefficients of the antenna pattern functions at the sidereal harmonics. Explicitly,
\begin{equation}
A^{+,\times}_k =
\frac{1}{T}
\int_0^T
F_{+,\times}(t,\psi=0)\,
e^{-i k \Omega_\oplus t}\, dt,
\qquad k=0,\pm1,\pm2.
\end{equation}
Using
\begin{equation}
H(t)=\Omega_\oplus t + (\beta-\alpha+\mathrm{GST}(t_0)),
\end{equation}
the antenna patterns can be written as
\begin{equation}
F_{+,\times}(t,0) =
\sum_{k=-2}^{2}
C^{+,\times}_k\,
e^{i k (\Omega_\oplus t + \beta-\alpha+\mathrm{GST}(t_0))},
\end{equation}
where $C_k^{+,\times}$ are real combinations of
$a_{kc},a_{ks},b_{kc},b_{ks}$.

If the constant phase $\mathrm{GST}(t_0)$ is absorbed into the reference time, the relation between the JKS coefficients and the 5-vector components reduces to
\begin{equation}
A^{+,\times}_{k,\mathrm{5vec}}
=
\Tilde{F}^{+,\times}(k\Omega_\oplus)\,
e^{i k (\alpha-\beta)},
\qquad k=0,\pm1,\pm2.
\end{equation}

The 5-vector and $\mathcal{F}$-statistic formalisms are mathematically equivalent descriptions of the same detector response to CWs. The apparent differences arise solely from different choices of time and phase parametrization. Once the sidereal phase conventions are aligned, the two approaches yield identical antenna patterns, likelihood functions, and detection statistics.

\subsection{Parameter degeneracies}
The complex 5-vector formalism makes the structure of parameter degeneracies particularly transparent. These degeneracies arise from the intrinsic symmetries of the CW signal and are summarized in Table \ref{tab:deg}.

\begin{figure}[t]
\centering
        \includegraphics[scale=0.6]{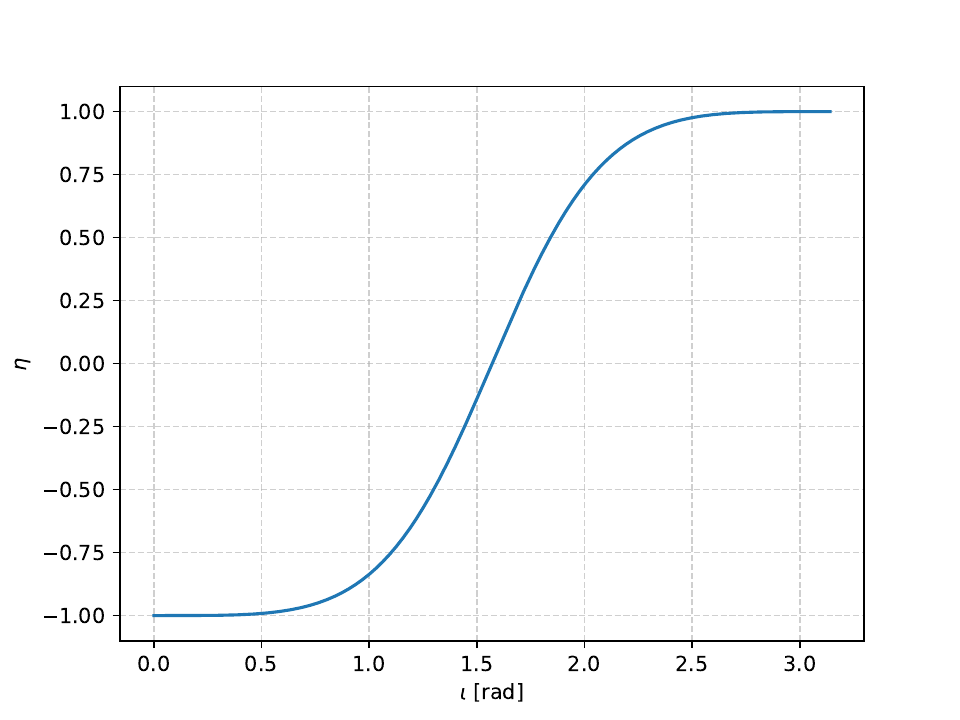}
    \caption{The polarization parameter $\eta$ in the 5-vector formalism as a function of the inclination angle $\iota$.}
    \vspace{-10pt}
    \label{fig:eta}
\end{figure}

The degree of degeneracy is controlled by the polarization parameter $\eta$, which depends on the inclination angle $\iota$ as shown in Fig.~\ref{fig:eta}. 

\begin{table}[t]
\centering
\begin{tabular}{c|c|c|l}
\hline
Case & $\eta$ & Polarization & Degeneracy structure \\
\hline

Linear & $\eta = 0$ &
linear &
\begin{tabular}{l}
$H_+ = \cos 2\psi$,\quad $H_\times = \sin 2\psi$ \\
$\psi \to \psi + \pi/2$,\ $\phi_0 \to \phi_0 + \pi$ \\
Sign flip degeneracy \\
\end{tabular}
\\

\hline

Circular (right) & $\eta = +1$ &
circular &
\begin{tabular}{l}
$H_\times = + j H_+$ \\
$\lambda_p \propto e^{j(\phi_0 - 2\psi)}$ \\
Only combination $(\phi_0 - 2\psi)$ measurable \\
Strong $\psi$–$\phi_0$ degeneracy \\
\end{tabular}
\\

\hline

Circular (left) & $\eta = -1$ &
circular &
\begin{tabular}{l}
$H_\times = - j H_+$ \\
$\lambda_p \propto e^{j(\phi_0 + 2\psi)}$ \\
Only combination $(\phi_0 + 2\psi)$ measurable \\
Strong $\psi$–$\phi_0$ degeneracy \\
\end{tabular}
\\
\end{tabular}
\caption{Polarization regimes and parameter degeneracies in the complex 5-vector CW formalism.}
\label{tab:deg}
\end{table}

\section{Heterodyned data comparison: 5-vectors computation}
In this appendix, we summarise the component-by-component comparison of the data 5-vector for the analysis of HI16 in O4a data among the three different data input: (i) the \texttt{py5vec} heterodyned data, obtained loading BSD data and reconstructing the phase evolution via \texttt{cwinpy}, (ii) the \texttt{SNAG} heterodyned data, obtained loading BSD data with an independent phase reconstruction in MATLAB, and (iii) the fully independent \texttt{HeterodynedData} product from \texttt{cwinpy}.

\begin{table}[h]
\centering 
\begin{tabular}{c|ccccc}
\hline
\texttt{py5vec} vs \texttt{SNAG} \\
$|R_i|$ & 1.036 & 1.023 & 1.067 & 1.038 & 1.045 \\
$\Delta \phi_i$ [$^\circ$] & -16.718 & -16.285 & -16.712 & -16.409 & -14.838 \\
\hline
\texttt{py5vec} vs \texttt{cwinpy} \\
$|R_i|$ & 1.019 & 1.041 & 1.050 & 1.048 & 1.013 \\
$\Delta \phi_i$ [$^\circ$] & -0.137 & 0.167 & -1.079 & -0.312 & -2.086 \\
\hline
\texttt{SNAG} vs \texttt{cwinpy} \\
$|R_i|$ & 0.984 & 1.017 & 0.984 & 1.010 & 0.970 \\
$\Delta \phi_i$ [$^\circ$] & 16.582 & 16.453 & 15.633 & 16.097 & 12.752 \\
\hline
\end{tabular}
\caption{Component-wise absolute ratios $|R_i|$ and phase differences $\Delta \phi_i$ for the five components of the data 5-vector between different pipelines.}
\label{tab:comp_5vec}
\end{table}

The comparison between \texttt{py5vec} and \texttt{cwinpy} shows excellent agreement, with ratios close to unity and phase differences below a few degrees for all components. The \texttt{py5vec} vs \texttt{SNAG} comparison instead reveals a nearly constant phase offset of about $-16^\circ$ across all five components. The \texttt{SNAG} vs \texttt{cwinpy} block confirms that this behaviour corresponds to a global phase rotation between \texttt{SNAG} and the other two pipelines, rather than to component-dependent discrepancies. Since \texttt{SNAG} and \texttt{py5vec} share the same BSD input format, the global phase offset must be due to the phase reconstruction. 

From the analysis of the other HI, the global offset does not depend on the detector but shows a dependence on the pulsar parameters. A more systematic analysis is needed to address the precise cause of the global phase offset.

\section{HI16 Bayesian parameters estimation}

This section completes the analysis of the hardware injection HI16 using O4a data from the two LIGO detectors described in Section \ref{sec:validation}.

\begin{figure}[t]
\centering
        \includegraphics[scale=0.43,
      trim=1cm 0cm 0cm 0cm,
      clip]{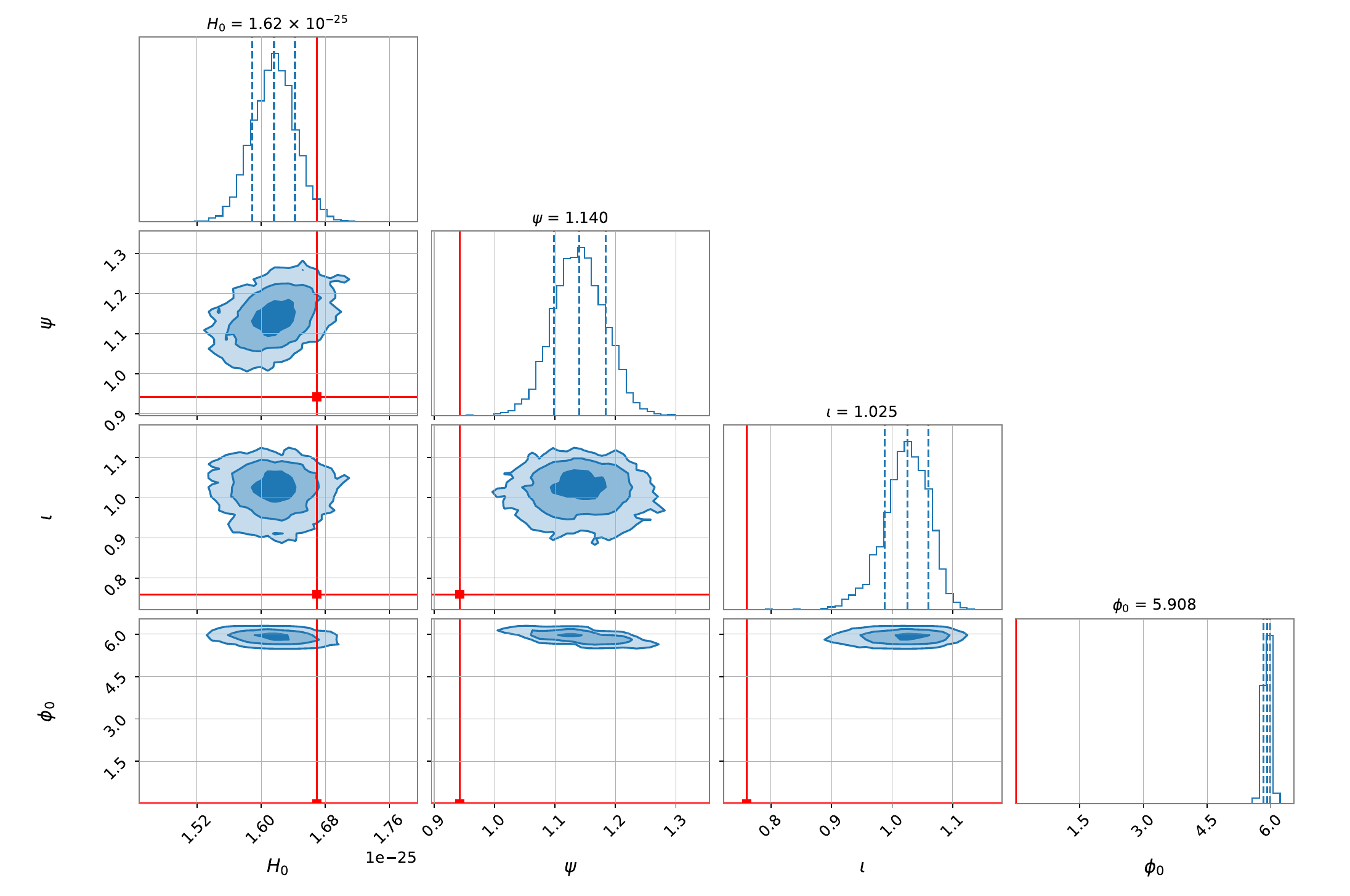}
    \caption{Posterior distribution functions for the signal parameters from the analysis of the hardware injection HI16 (binary system) considering the O4a dataset of the LIGO Hanford detector. The red dots/lines show the injected value. For this plot, we take as input data for \texttt{py5vec}, the \texttt{HeterodynedData} from \texttt{cwinpy}, and we use the Gaussian likelihood in Eq.\eqref{likelihood_MD}.}
    \vspace{-10pt}
    \label{fig:het_data_HI16_O4a_PE}
\end{figure}

Figure \ref{fig:het_data_HI16_O4a_PE} shows the results obtained with \texttt{py5vec}, using as input the $\texttt{HeterodynedData}$ produced by \texttt{cwinpy}. We adopt the Gaussian likelihood defined in Eq.\eqref{likelihood_MD} and used for Figure \ref{fig:HI16_O4a_PE}. The posteriors for the LIGO Hanford O4a data are in agreement with those shown in Figure \ref{fig:HI16_O4a_PE}. This confirms that the shift of the narrow posteriors with respect to the injected values does not depend on the input data or on the applied correction but rather on the adopted likelihood.

\begin{figure}[t]
\centering
        \includegraphics[scale=0.5,
      trim=0cm 0cm 0cm 0cm,
      clip]{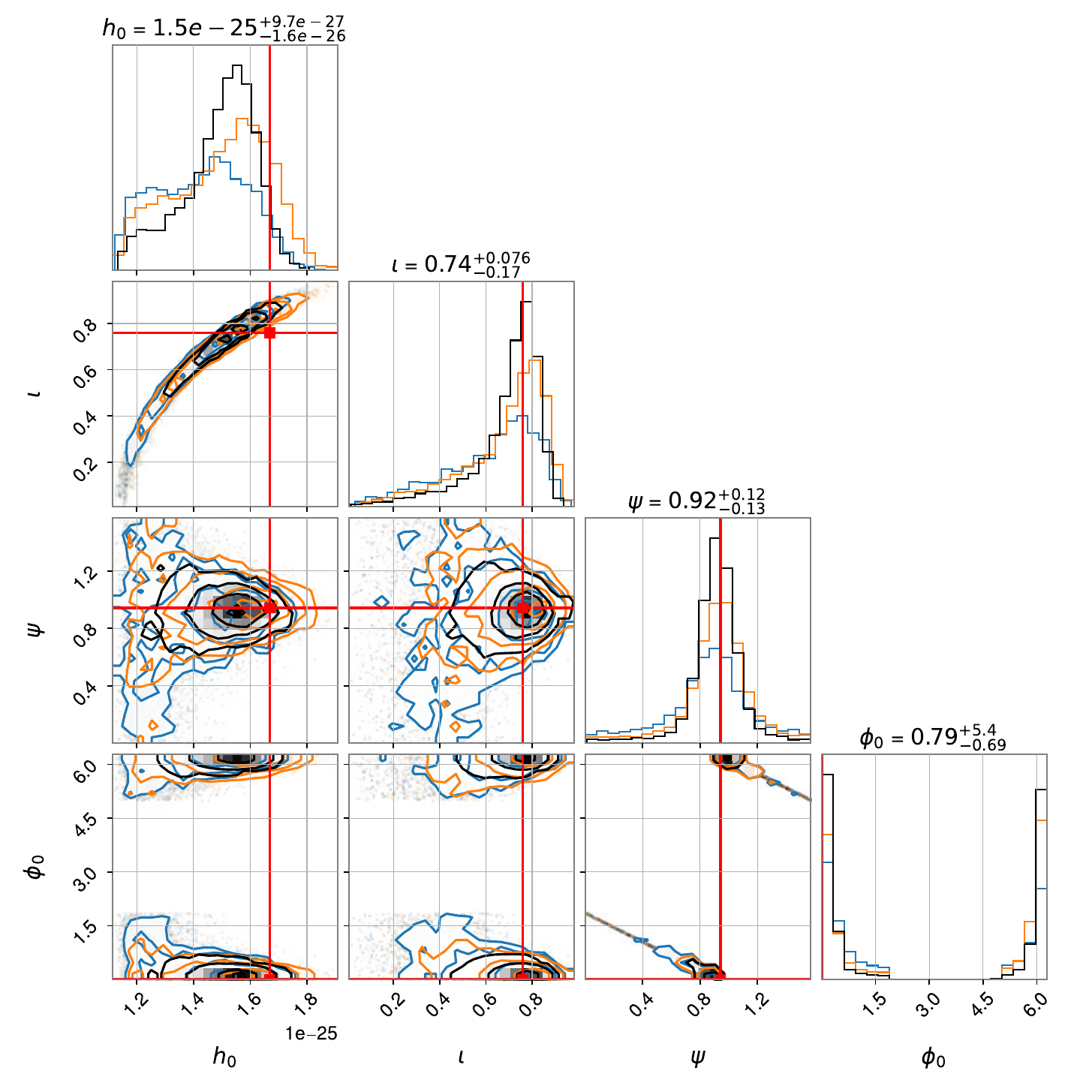}
    \caption{\texttt{cwinpy} posterior distribution functions for the signal parameters from the analysis of the hardware injection HI16 considering the O4a datasets of the LIGO Livingstone (in blue), LIGO Hanford (in orange), and the multidetector. The red dots/lines show the injected value. Note that the amplitude posterior is in $h_0$ and not in $H_0$ (see Eq.\eqref{corramp}). The used Student's t-likelihood is defined in Eq.12 of \cite{bayesian}.}
    \vspace{-10pt}
    \label{fig:cwinpy_HI16_O4a_PE}
\end{figure}

Figure~\ref{fig:cwinpy_HI16_O4a_PE} shows the \texttt{cwinpy} results for the analysis of HI16 using the O4a LIGO data. In this case, the likelihood is the Student's $t$-likelihood, as defined in \cite{bayesian}. The posteriors show very good agreement with the results obtained using \texttt{py5vec} with a conceptually similar likelihood. The main difference appears in the amplitude parameter: \texttt{cwinpy} samples in $h_0$, which shows a correlation with the inclination angle $\iota$, while in our analysis the posterior is expressed in terms of $H_0$, whose definition (see Eq. \eqref{corramp}) absorbs the $\cos\iota$ dependence and therefore partially masks this correlation.

\singlespacing
\printbibliography 

@PREAMBLE{
 "\providecommand{\noopsort}[1]{}" 
 # "\providecommand{\singleletter}[1]{#1}%" 
}

@article{Amicucci_2025,
doi = {10.1088/1361-6382/adecd7},
url = {https://doi.org/10.1088/1361-6382/adecd7},
year = {2025},
month = {jul},
publisher = {IOP Publishing},
volume = {42},
number = {14},
pages = {145008},
author = {Amicucci, Francesco and Leaci, Paola and Astone, Pia and D’Antonio, Sabrina and Dal Pra, Stefano and Di Giovanni, Matteo and D’Onofrio, Luca and Muciaccia, Federico and Palomba, Cristiano and Pierini, Lorenzo and Singhal, Akshat},
title = {A directed continuous-wave search from Scorpius X-1 with the five-vector resampling technique},
journal = {Classical and Quantum Gravity}
}

@article{spinwand,
  title = {How much spin wandering can continuous gravitational wave search algorithms handle?},
  author = {Carlin, Julian B. and Melatos, Andrew},
  journal = {Phys. Rev. D},
  volume = {111},
  issue = {8},
  pages = {083016},
  numpages = {16},
  year = {2025},
  month = {Apr},
  publisher = {American Physical Society},
  doi = {10.1103/PhysRevD.111.083016},
  url = {https://link.aps.org/doi/10.1103/PhysRevD.111.083016}
}

@misc{Snag,
  author       = {Sergio Frasca},
  title        = {Snag},
  year         = {2026},
  howpublished = {MATLAB Central File Exchange},
  url          = {https://it.mathworks.com/matlabcentral/fileexchange/9703-snag}
}

@article{folding,
  title = {Fast gravitational wave radiometry using data folding},
  author = {Ain, Anirban and Dalvi, Prathamesh and Mitra, Sanjit},
  journal = {Phys. Rev. D},
  volume = {92},
  issue = {2},
  pages = {022003},
  numpages = {12},
  year = {2015},
  month = {Jul},
  publisher = {American Physical Society},
  doi = {10.1103/PhysRevD.92.022003},
  url = {https://link.aps.org/doi/10.1103/PhysRevD.92.022003}
}

@misc{O4HI,
      title={Monitoring of Continuous-Wave Hardware Injections in LIGO Interferometers during the O4 Observing Run}, 
      author={Preet Baxi and Jessica Leviton and Eilam Morag and Matthew Pitkin and Keith Riles},
      year={2026},
      eprint={2601.09918},
      archivePrefix={arXiv},
      primaryClass={astro-ph.IM},
      url={https://arxiv.org/abs/2601.09918}, 
}

@article{marg_phi0,
doi = {10.1088/0264-9381/31/6/065002},
url = {https://doi.org/10.1088/0264-9381/31/6/065002},
year = {2014},
month = {feb},
publisher = {IOP Publishing},
volume = {31},
number = {6},
pages = {065002},
author = {Whelan, John T and Prix, Reinhard and Cutler, Curt J and Willis, Joshua L},
title = {New coordinates for the amplitude parameter space of continuous gravitational waves},
journal = {Classical and Quantum Gravity},
}

@article{Jones2015ParameterCA,
  title={Parameter choices and ranges for continuous gravitational wave searches for steadily spinning neutron stars},
  author={David A. Jones},
  journal={Monthly Notices of the Royal Astronomical Society},
  year={2015},
  volume={453},
  pages={53-66},
  url={https://api.semanticscholar.org/CorpusID:33854752}
}

@article{bilby_paper,
    author = "Ashton, Gregory and others",
    title = "{BILBY: A user-friendly Bayesian inference library for gravitational-wave astronomy}",
    eprint = "1811.02042",
    archivePrefix = "arXiv",
    primaryClass = "astro-ph.IM",
    doi = "10.3847/1538-4365/ab06fc",
    journal = "Astrophys. J. Suppl.",
    volume = "241",
    number = "2",
    pages = "27",
    year = "2019"
}

@article{PhysRevD.96.063004,
  title = {Statistical characterization of pulsar glitches and their potential impact on searches for continuous gravitational waves},
  author = {Ashton, G. and Prix, R. and Jones, D. I.},
  journal = {Phys. Rev. D},
  volume = {96},
  issue = {6},
  pages = {063004},
  numpages = {23},
  year = {2017},
  month = {Sep},
  publisher = {American Physical Society},
  doi = {10.1103/PhysRevD.96.063004},
  url = {https://link.aps.org/doi/10.1103/PhysRevD.96.063004}
}

@article{O4a,
doi = {10.3847/1538-4357/adb3a0},
url = {https://doi.org/10.3847/1538-4357/adb3a0},
year = {2025},
month = {apr},
publisher = {The American Astronomical Society},
volume = {983},
number = {2},
pages = {99},
author = {Abac, A. G. and Abbott, R. and Abouelfettouh, I. and Acernese, F. and Ackley, K. and Adhicary, S. and Adhikari, N. and Adhikari, R. X. and Adkins, V. K. and Agarwal, D. and Agathos, M. and Aghaei Abchouyeh, M. and Aguiar},
title = {Search for Continuous Gravitational Waves from Known Pulsars in the First Part of the Fourth LIGO-Virgo-KAGRA Observing Run},
journal = {The Astrophysical Journal},
}

@article{Prix_2025,
doi = {10.1088/1361-6382/adb097},
url = {https://doi.org/10.1088/1361-6382/adb097},
year = {2025},
month = {feb},
publisher = {IOP Publishing},
volume = {42},
number = {6},
pages = {065006},
author = {Prix, Reinhard},
title = {Analytic weak-signal approximation of the Bayes factor for continuous gravitational waves},
journal = {Classical and Quantum Gravity},
}

@article{Ashok,
  title = {Bayesian $\mathcal{F}$-statistic-based parameter estimation of continuous gravitational waves from known pulsars},
  author = {Ashok, A. and Covas, P. B. and Prix, R. and Papa, M. A.},
  journal = {Phys. Rev. D},
  volume = {109},
  issue = {10},
  pages = {104002},
  numpages = {15},
  year = {2024},
  month = {May},
  publisher = {American Physical Society},
  doi = {10.1103/PhysRevD.109.104002},
  url = {https://link.aps.org/doi/10.1103/PhysRevD.109.104002}
}

@article{Mirasola,
  title = {Toward a computationally efficient follow-up pipeline for blind continuous gravitational-wave searches},
  author = {Mirasola, Lorenzo and Tenorio, Rodrigo},
  journal = {Phys. Rev. D},
  volume = {110},
  issue = {12},
  pages = {124049},
  numpages = {19},
  year = {2024},
  month = {Dec},
  publisher = {American Physical Society},
  doi = {10.1103/PhysRevD.110.124049},
  url = {https://link.aps.org/doi/10.1103/PhysRevD.110.124049}
}

@ARTICLE{2023LRR....26....3R,
       author = {{Riles}, Keith},
        title = "{Searches for continuous-wave gravitational radiation}",
      journal = {Living Reviews in Relativity},
         year = 2023,
        month = dec,
       volume = {26},
       number = {1},
          eid = {3},
        pages = {3},
          doi = {10.1007/s41114-023-00044-3},
archivePrefix = {arXiv},
       eprint = {2206.06447},
 primaryClass = {astro-ph.HE},
       adsurl = {https://ui.adsabs.harvard.edu/abs/2023LRR....26....3R}
}

@article{WETTE2023102880,
title = {Searches for continuous gravitational waves from neutron stars: A twenty-year retrospective},
journal = {Astroparticle Physics},
volume = {153},
pages = {102880},
year = {2023},
issn = {0927-6505},
doi = {https://doi.org/10.1016/j.astropartphys.2023.102880},
url = {https://www.sciencedirect.com/science/article/pii/S092765052300066X},
author = {Karl Wette}
}

@article{Pitkin2022, 
    doi = {10.21105/joss.04568}, 
    url = {https://doi.org/10.21105/joss.04568}, 
    year = {2022}, 
    publisher = {The Open Journal}, 
    volume = {7}, 
    number = {77}, 
    pages = {4568}, 
    author = {Matthew Pitkin}, 
    title = {CWInPy: A Python package for inference with continuous gravitational-wave signals from pulsars}, 
    journal = {Journal of Open Source Software}
}

@article{DOnofrio_2025,
doi = {10.1088/1361-6382/ad94c5},
url = {https://dx.doi.org/10.1088/1361-6382/ad94c5},
year = {2024},
month = {dec},
publisher = {IOP Publishing},
volume = {42},
number = {1},
pages = {015005},
author = {L D’Onofrio and P Astone and S Dal Pra and S D’Antonio and M Di Giovanni and R De Rosa and P Leaci and S Mastrogiovanni and L Mirasola and F Muciaccia and C Palomba and L Pierini},
title = {Two sides of the same coin: the $\mathcal{F}$-statistic and the 5-vector method},
journal = {Classical and Quantum Gravity}
}

@misc{O4,
  author = {{LVK Collaboration}},
  title = {{Gravitational-Wave Observatory Status}},
  howpublished = "\url{https://gwosc.org/detector_status/}",
  year = {2024}
}

@article{Fbayesnew,
  title = {Bayesian $\mathcal{F}$-statistic-based parameter estimation of continuous gravitational waves from known pulsars},
  author = {Ashok, A. and Covas, P. B. and Prix, R. and Papa, M. A.},
  journal = {Phys. Rev. D},
  volume = {109},
  issue = {10},
  pages = {104002},
  numpages = {15},
  year = {2024},
  month = {May},
  publisher = {American Physical Society},
  doi = {10.1103/PhysRevD.109.104002},
  url = {https://link.aps.org/doi/10.1103/PhysRevD.109.104002}
}

@article{semicoh,
  title = {Semicoherent method to search for continuous gravitational waves},
  author = {D'Antonio, Sabrina and Palomba, Cristiano and Astone, Pia and Dall'Osso, Simone and Dal Pra, Stefano and Frasca, Sergio and Leaci, Paola and Muciaccia, Federico and Piccinni, Ornella J. and Pierini, Lorenzo and Serra, Marco},
  journal = {Phys. Rev. D},
  volume = {108},
  issue = {12},
  pages = {122001},
  numpages = {16},
  year = {2023},
  month = {Dec},
  publisher = {American Physical Society},
  doi = {10.1103/PhysRevD.108.122001},
  url = {https://link.aps.org/doi/10.1103/PhysRevD.108.122001}
}

@ARTICLE{ligo,
       author = {{LIGO Scientific Collaboration}},
        title = "{Advanced LIGO}",
      journal = {Classical and Quantum Gravity},
         year = {2015},
        month = {04},
       volume = {32},
       number = {7},
          eid = {074001},
        pages = {074001},
          doi = {10.1088/0264-9381/32/7/074001},
archivePrefix = {arXiv},
       eprint = {1411.4547},
 primaryClass = {gr-qc},
       adsurl = {https://ui.adsabs.harvard.edu/abs/2015CQGra..32g4001L}
}

@ARTICLE{virgo,
       author = {{F. Acernese} et al},
        title = "{Advanced Virgo: a second-generation interferometric gravitational wave detector}",
      journal = {Classical and Quantum Gravity},
         year = {2015},
        month = {01},
       volume = {32},
       number = {2},
          eid = {024001},
        pages = {024001},
          doi = {10.1088/0264-9381/32/2/024001},
archivePrefix = {arXiv},
       eprint = {1408.3978},
 primaryClass = {gr-qc},
       adsurl = {https://ui.adsabs.harvard.edu/abs/2015CQGra..32b4001A}
}

@article{2014,
  title = {Method for narrow-band search of continuous gravitational wave signals},
  author = {Astone, P. and Colla, A. and D'Antonio, S. and Frasca, S. and Palomba, C. and Serafinelli, R.},
  journal = {Phys. Rev. D},
  volume = {89},
  issue = {6},
  pages = {062008},
  numpages = {11},
  year = {2014},
  month = {Mar},
  publisher = {American Physical Society},
  doi = {10.1103/PhysRevD.89.062008},
  url = {https://link.aps.org/doi/10.1103/PhysRevD.89.062008}
}

@article{JKS,
  title = {{Data analysis of gravitational-wave signals from spinning neutron stars: The signal and its detection}},
  author = {Jaranowski, Piotr and Kr\'olak, Andrzej and Schutz, Bernard F.},
  journal = {Phys. Rev. D},
  volume = {58},
  issue = {6},
  pages = {063001},
  numpages = {24},
  year = {1998},
  month = {Aug},
  publisher = {American Physical Society},
  doi = {10.1103/PhysRevD.58.063001},
  url = {https://link.aps.org/doi/10.1103/PhysRevD.58.063001}
}

@article{mio,
author = {Buono, Mauro and De Rosa, Rosario and D'Onofrio, Luca and Errico, Luciano and Palomba, Cristiano and Piccinni, Ornella and Sequino, Valeria},
year = {2021},
month = {03},
title = {A method for detecting continuous gravitational wave signals from an ensemble of known pulsars},
volume = {38},
journal = {Classical and Quantum Gravity},
doi = {10.1088/1361-6382/abf1c0}
}

@article{2010,
author = {Astone, P. and D'Antonio, S. and Frasca, S and Palomba, C},
year = {2010},
month = {09},
pages = {194016},
title = {A method for detection of known sources of continuous gravitational wave signals in non-stationary data},
volume = {27},
journal = {Classical and Quantum Gravity},
doi = {10.1088/0264-9381/27/19/194016}
}

@article{mio2,
    author = "D'Onofrio, Luca and De Rosa, Rosario and Errico, Luciano and Palomba, Cristiano and Sequino, Valeria and Trozzo, Lucia",
    title = "{5n-vector ensemble method for detecting gravitational waves from known pulsars}",
    doi = "10.1103/PhysRevD.105.063012",
    journal = "Phys. Rev. D",
    volume = "105",
    number = "6",
    pages = "063012",
    year = "2022"
}

@article{ensbayes,
    author = "Pitkin, M. and Messenger, C. and Fan, X.",
    title = "{Hierarchical Bayesian method for detecting continuous gravitational waves from an ensemble of pulsars}",
    primaryClass = "astro-ph.IM",
    reportNumber = "LIGO-P1800171, INT-PUB-18-037",
    doi = "10.1103/PhysRevD.98.063001",
    journal = "Phys. Rev. D",
    volume = "98",
    number = "6",
    pages = "063001",
    year = "2018"
}

@article{2019,
  title={A new data analysis framework for the search of continuous gravitational wave signals},
  author={Ornella Juliana Piccinni and Pia Astone and Sabrina D’Antonio and Sergio Frasca and Giuseppe Intini and Paola Leaci and S Mastrogiovanni and A. L. Miller and Cristiano Palomba and Akshat Singhal},
  journal={Classical and Quantum Gravity},
  year={2018}
}

@article{HI,
  title = {Validating gravitational-wave detections: The Advanced LIGO hardware injection system},
  author = {Biwer, C. and Barker, D. and Batch, J. C. and Betzwieser, J. and Fisher, R. P. and Goetz, E. and Kandhasamy, S. and Karki, S. and Kissel, J. S. and Lundgren, A. P. and Macleod, D. M. and Mullavey, A. and Riles, K. and Rollins, J. G. and Thorne, K. A. and Thrane, E. and Abbott, T. D. and Allen, B. and Brown, D. A. and Charlton, P. and Crowder, S. G. and Fritschel, P. and Kanner, J. B. and Landry, M. and Lazzaro, C. and Millhouse, M. and Pitkin, M. and Savage, R. L. and Shawhan, P. and Shoemaker, D. H. and Smith, J. R. and Sun, L. and Veitch, J. and Vitale, S. and Weinstein, A. J. and Cornish, N. and Essick, R. C. and Fays, M. and Katsavounidis, E. and Lange, J. and Littenberg, T. B. and Lynch, R. and Meyers, P. M. and Pannarale, F. and Prix, R. and O'Shaughnessy, R. and Sigg, D.},
  journal = {Phys. Rev. D},
  volume = {95},
  issue = {6},
  pages = {062002},
  numpages = {15},
  year = {2017},
  month = {Mar},
  publisher = {American Physical Society},
  doi = {10.1103/PhysRevD.95.062002},
  url = {https://link.aps.org/doi/10.1103/PhysRevD.95.062002}
}

@article{eterodina,
  title = {Bayesian estimation of pulsar parameters from gravitational wave data},
  author = {Dupuis, R\'ejean J. and Woan, Graham},
  journal = {Phys. Rev. D},
  volume = {72},
  issue = {10},
  pages = {102002},
  numpages = {9},
  year = {2005},
  month = {Nov},
  publisher = {American Physical Society},
  doi = {10.1103/PhysRevD.72.102002},
  url = {https://link.aps.org/doi/10.1103/PhysRevD.72.102002}
}

@ARTICLE{ornella,
       author = {{Ornella Juliana}, Piccinni},
        title = "{Status and perspectives of Continuous Gravitational Wave searches}",
      journal = {arXiv e-prints},
         year = 2022,
        month = feb,
          eid = {arXiv:2202.01088},
        pages = {arXiv:2202.01088},
archivePrefix = {arXiv},
       eprint = {2202.01088},
 primaryClass = {gr-qc},
       adsurl = {https://ui.adsabs.harvard.edu/abs/2022arXiv220201088O}
}

@ARTICLE{narrowband,
	author={The LIGO and Virgo Collaboration},
        title = "{Narrowband Searches for Continuous and Long-duration Transient Gravitational Waves from Known Pulsars in the LIGO-Virgo Third Observing Run}",
    journal = {The Astrophysical Journal},
         year = 2022,
        month = jun,
       volume = {932},
       number = {2},
          eid = {133},
        pages = {133},
          doi = {10.3847/1538-4357/ac6ad0},
       adsurl = {https://ui.adsabs.harvard.edu/abs/2022ApJ...932..133A}
}

@online{bayesian,
  doi = {10.48550/ARXIV.1705.08978},
  
  url = {https://arxiv.org/abs/1705.08978},
  
  author = {Pitkin, Matthew and Isi, Maximiliano and Veitch, John and Woan, Graham},
  
  title = {A nested sampling code for targeted searches for continuous gravitational waves from pulsars},
  
  publisher = {arXiv},
  
  year = {2017},
  
  copyright = {arXiv.org perpetual, non-exclusive license}
}

@misc{cwemission,
    author = "Jones, D. I.",
    title = "{Learning from the Frequency Content of Continuous Gravitational Wave Signals}",
    eprint = "2111.08561",
    archivePrefix = "arXiv",
    primaryClass = "astro-ph.HE",
    month = "11",
    year = "2021"
}

@article{mio3,
  title = {Search for gravitational wave signals from known pulsars in LIGO-Virgo O3 data using the $5n$-vector ensemble method},
  author = {D'Onofrio, Luca and De Rosa, Rosario and Palomba, Cristiano and Leaci, Paola and Piccinni, Ornella J. and Sequino, Valeria and Errico, Luciano and Trozzo, Lucia and Palfreyman, Jim and McKee, James W. and Meyers, Bradley W. and Stairs, Ingrid and Guillemot, Lucas and Cognard, Isma\"el and Theureau, Gilles and Keith, Michael J. and Lyne, Andrew and Flynn, Chris and Stappers, Ben},
  journal = {Phys. Rev. D},
  volume = {108},
  issue = {12},
  pages = {122002},
  numpages = {20},
  year = {2023},
  month = {Dec},
  publisher = {American Physical Society},
  doi = {10.1103/PhysRevD.108.122002},
  url = {https://link.aps.org/doi/10.1103/PhysRevD.108.122002}
}

@article{KAGRA:2020tym,
    author = "Akutsu, T. and others",
    collaboration = "KAGRA",
    title = "{Overview of KAGRA: Detector design and construction history}",
    eprint = "2005.05574",
    archivePrefix = "arXiv",
    primaryClass = "physics.ins-det",
    doi = "10.1093/ptep/ptaa125",
    journal = "PTEP",
    volume = "2021",
    number = "5",
    pages = "05A101",
    year = "2021"
}

@article{LIGOScientific:2025snk,
    author = "Abac, A. G. and others",
    collaboration = "LIGO Scientific, VIRGO, KAGRA",
    title = "{Open Data from LIGO, Virgo, and KAGRA through the First Part of the Fourth Observing Run}",
    eprint = "2508.18079",
    archivePrefix = "arXiv",
    primaryClass = "gr-qc",
    reportNumber = "LIGO-P2500167",
    month = "8",
    year = "2025"
}

@article{Baxi:2026wbr,
    author = "Baxi, Preet and Leviton, Jessica and Morag, Eilam and Pitkin, Matthew and Riles, Keith",
    title = "{Monitoring of Continuous-Wave Hardware Injections in LIGO Interferometers during the O4 Observing Run}",
    eprint = "2601.09918",
    archivePrefix = "arXiv",
    primaryClass = "astro-ph.IM",
    month = "1",
    year = "2026"
}

@article{Keitel:2021xeq,
    author = "Keitel, David and Tenorio, Rodrigo and Ashton, Gregory and Prix, Reinhard",
    title = "{PyFstat: a Python package for continuous gravitational-wave data analysis}",
    eprint = "2101.10915",
    archivePrefix = "arXiv",
    primaryClass = "gr-qc",
    reportNumber = "LIGO-P2100008",
    doi = "10.21105/joss.03000",
    journal = "J. Open Source Softw.",
    volume = "6",
    number = "60",
    pages = "3000",
    year = "2021"
}

@article{PhysRevD.97.103020,
    author = "Ashton, Gregory and Prix, Reinhard",
    title = "{Hierarchical multistage MCMC follow-up of continuous gravitational wave candidates}",
    eprint = "1802.05450",
    archivePrefix = "arXiv",
    primaryClass = "astro-ph.IM",
    doi = "10.1103/PhysRevD.97.103020",
    journal = "Phys. Rev. D",
    volume = "97",
    number = "10",
    pages = "103020",
    year = "2018"
}

@misc{lalsuite,
       author         = "{LIGO Scientific Collaboration} and {Virgo Collaboration} and {KAGRA Collaboration}",
       title          = "{LVK} {A}lgorithm {L}ibrary - {LALS}uite",
       howpublished   = "Free software (GPL)",
       doi            = "10.7935/GT1W-FZ16",
       year           = "2018"
 }

@article{swiglal,
          title     = "{SWIGLAL: Python and Octave interfaces to the LALSuite gravitational-wave data analysis libraries}",
          author    = "Karl Wette",
          journal   = "SoftwareX",
          volume    = "12",
          pages     = "100634",
          year      = "2020",
          doi       = "10.1016/j.softx.2020.100634"
 }

@article{Speagle:2019ivv,
    author = "Speagle, Joshua S.",
    title = "{dynesty: a dynamic nested sampling package for estimating Bayesian posteriors and evidences}",
    eprint = "1904.02180",
    archivePrefix = "arXiv",
    primaryClass = "astro-ph.IM",
    doi = "10.1093/mnras/staa278",
    journal = "Mon. Not. Roy. Astron. Soc.",
    volume = "493",
    number = "3",
    pages = "3132--3158",
    year = "2020"
}

@ARTICLE{2020SciPy-NMeth,
  author  = {Virtanen, Pauli and Gommers, Ralf and Oliphant, Travis E. and
            Haberland, Matt and Reddy, Tyler and Cournapeau, David and
            Burovski, Evgeni and Peterson, Pearu and Weckesser, Warren and
            Bright, Jonathan and {van der Walt}, St{\'e}fan J. and
            Brett, Matthew and Wilson, Joshua and Millman, K. Jarrod and
            Mayorov, Nikolay and Nelson, Andrew R. J. and Jones, Eric and
            Kern, Robert and Larson, Eric and Carey, C J and
            Polat, {\.I}lhan and Feng, Yu and Moore, Eric W. and
            {VanderPlas}, Jake and Laxalde, Denis and Perktold, Josef and
            Cimrman, Robert and Henriksen, Ian and Quintero, E. A. and
            Harris, Charles R. and Archibald, Anne M. and
            Ribeiro, Ant{\^o}nio H. and Pedregosa, Fabian and
            {van Mulbregt}, Paul and {SciPy 1.0 Contributors}},
  title   = {{{SciPy} 1.0: Fundamental Algorithms for Scientific
            Computing in Python}},
  journal = {Nature Methods},
  year    = {2020},
  volume  = {17},
  pages   = {261--272},
  adsurl  = {https://rdcu.be/b08Wh},
  doi     = {10.1038/s41592-019-0686-2},
}

\end{document}